\documentclass[a4paper,11pt]{article}
\pdfoutput=1 

\usepackage{jheppub} 

\usepackage{amsmath}
\usepackage[svgnames]{xcolor}
\usepackage{physics}
\usepackage{amsmath}
\usepackage{mathdots}
\usepackage{yhmath}
\usepackage{cancel}
\usepackage{color}
\usepackage{siunitx}
\usepackage{array}
\usepackage{multirow}
\usepackage{amssymb}
\usepackage{gensymb}
\usepackage{tabularx}
\usepackage{extarrows}
\usepackage{booktabs}
\usepackage{bm}
\usepackage{comment}

\usepackage{tikz}
\usepackage{mathdots}
\usepackage{yhmath}
\usepackage{cancel}
\usepackage{color}
\usepackage{siunitx}
\usepackage{array}
\usepackage{multirow}
\usepackage{amssymb}
\usepackage{gensymb}
\usepackage{tabularx}
\usepackage{extarrows}
\usepackage{booktabs}
\usetikzlibrary{fadings}
\usetikzlibrary{patterns}
\usetikzlibrary{shadows.blur}
\usetikzlibrary{shapes}
\allowdisplaybreaks

\makeatletter
\def\@fpheader{~}
\makeatother

\title{Transverse momentum dependent distribution functions in the threshold limit}

\author[a,b,c]{Zhong-Bo Kang,}
\author[a,d]{Kajal Samanta,}
\author[d,e,f]{Ding Yu Shao}
\author[d]{and Yang-Li Zeng}

\affiliation[a]{Department of Physics and Astronomy, University of California, Los Angeles, CA 90095, USA}
\affiliation[b]{Mani L. Bhaumik Institute for Theoretical Physics, University of California, Los Angeles, CA 90095, USA}
\affiliation[c]{Center for Frontiers in Nuclear Science, Stony Brook University, Stony Brook, NY 11794, USA}
\affiliation[d]{Department of Physics and Center for Field Theory and Particle Physics, Fudan University, Shanghai 200438, China}
\affiliation[e]{Key Laboratory of Nuclear Physics and Ion-beam Application (MOE), Fudan University,
Shanghai 200433, China}
\affiliation[f]{Shanghai Research Center for Theoretical Nuclear Physics, NSFC and Fudan University, Shanghai 200438, China}

\emailAdd{zkang@ucla.edu}
\emailAdd{kajal003@ucla.edu}
\emailAdd{dingyu.shao@cern.ch}
\emailAdd{ylzeng19@fudan.edu.cn}

\abstract{We apply the joint threshold and transverse momentum dependent (TMD) factorization theorem to introduce new threshold-TMD distribution functions, including threshold-TMD parton distribution functions (PDFs) and fragmentation functions (FFs). We apply Soft-Collinear Effective Theory and renormalization group methods to carry out QCD evolution for both threshold-TMD PDFs and FFs. We show the universality of threshold-TMD functions among three standard processes, i.e. the Drell-Yan production in $pp$ collisions, semi-inclusive deep-inelastic scattering and back-to-back two hadron production in $e^+e^-$ collisions. In the end, we present the numerical predictions for different threshold-TMD functions and also transverse momentum distributions at $pp$, $ep$, and $e^+e^-$ collisions. }

\begin{document} 
\maketitle
\flushbottom

\section{Introduction}

The parton distribution functions (PDFs) and fragmentation functions (FFs) are two important quantities in particle physics to understand the dynamics of a parton inside a hadron. These functions have been studied extensively both by experiment and theory hadron physics community~\cite{Lin:2017snn}. In the last decade, the hadron physics community proposed a new kind of distribution function called transverse momentum dependent (TMD) distribution functions to extend this studies~\cite{Accardi:2012qut,AbdulKhalek:2021gbh}. These TMD functions provide information on a parton carrying a certain amount of longitudinal momentum fraction $x$ and transverse momentum $\bm k_T$ inside a nucleon, which can be used to probe the quantum correlation between the nucleon spin and active quark or gluon polarization as well as its motion. The measurement of TMD observables provides the leading information on the three-dimensional imaging of a nucleon. 

The TMD factorization and resummation framework offer a bridge between TMD functions and observations, which was originally obtained by Collins, Soper and Sterman in \cite{Collins:1981uk,Collins:1984kg} and has also been derived in Soft-Collinear Effective Theory (SCET)~\cite{Bauer:2000yr,Bauer:2001ct,Bauer:2001yt,Bauer:2002nz,Beneke:2002ph} based on renormalization group (RG) methods \cite{Becher:2010tm,Chiu:2011qc,GarciaEchevarria:2011rb}. In the small transverse momentum limit, the differential cross section can be factorized as the product of the hard factor and TMD functions at the leading power. Therefore, one can directly probe  TMD functions via different processes, including the Drell-Yan, semi-inclusive deep-inelastic scattering (SIDIS) and back-to-back two hadron production in $e^+e^-$ collisions. The universality of the non-perturbative parametrization of TMD functions has been investigated extensively in~\cite{Su:2014wpa,Boglione:2016bph,Hautmann:2020cyp,Collins:2004nx,Kang:2015msa,Bacchetta:2017gcc,Scimemi:2019cmh,Cammarota:2020qcw,Bacchetta:2020gko,Echevarria:2020hpy,Bury:2022czx,Bacchetta:2022awv}. In the context of the small-$x$ limit, the naive TMD factorization formula at the leading-power might no longer apply, and numerous studies have addressed TMD functions with considerations for gluon saturation effects \cite{Balitsky:2015qba, Zhou:2016tfe, Xiao:2017yya, Zhou:2018lfq, Boer:2022njw}. Conversely, as $x$ approaches larger values, the underlying factorization theorem of leading-power TMD factorization remains robust; nonetheless, the associated resummation formula begins to incorporate non-negligible logarithmic contributions. Intriguingly, recent lattice QCD simulations \cite{LPC:2022zci} demonstrate that in this large $x$ regime, lattice results show inconsistencies with the predictions of various existing non-perturbative parametrization of TMD PDF models \cite{Bacchetta:2017gcc,Scimemi:2019cmh,Bury:2022czx,Bacchetta:2022awv}. Extending our comprehension of TMD functions in this limit is thus of pivotal importance. This study aims to extend our understanding of TMD functions in the large $x$ limit.

In the threshold limit, the phase space for real radiation is restricted, and then the infrared cancellations between real and virtual diagrams leave behind large logarithms $\sim \ln(1-x)$. Therefore, near the threshold limit, it becomes necessary to take into account these large logarithmic corrections to all orders to have a reliable theoretical prediction. In the Mellin space, these large logarithms are transformed into powers of logarithms of the Mellin variable $N$. A systematic approach has been proposed to resum these large logarithms to all order \cite{Sterman:1986aj,Catani:1989ne} in the Mellin space and the technique is known as threshold resummation. It is noted that, unlike the TMD resummation, the singular threshold logarithms do not appear explicitly in the physical cross section, since they are always convoluted with PDFs or FFs at the hadron level. After analyzing the dynamical origins of the large corrections in both threshold and TMD resummations, the joint resummation framework of threshold and TMD logarithms was first developed in \cite{Laenen:2000de}. Such a framework has been applied in various processes \cite{Kulesza:2002rh,Kulesza:2003wn,Banfi:2004xa,Bozzi:2007tea,Debove:2011xj} at hadron colliders. Later on, a factorization and resummation formalism based on SCET$_+$ \cite{Bauer:2011uc, Procura:2014cba} was introduced in \cite{Lustermans:2016nvk,Li:2016axz}, which can be used to perform resummation calculation beyond the next-to-leading logarithmic (NLL) accuracy.

In this paper, we introduce a new type of unpolarized TMD functions, threshold-TMD distribution functions, within the joint threshold and TMD factorization and resummation framework. We apply the crossed threshold resummation method \cite{Sterman:2006hu} to find a close correspondence between resummation for Drell-Yan, SIDIS and $e^+e^-$ processes. Therefore, one can obtain the resummation formula for each of the processes using the same procedure. In the joint limit, the cross section is factorized as the product of hard factor and threshold-TMD functions, including threshold-TMD PDFs and FFs, and the structure of the logarithms turns out to be identical.  Among these three processes, we have the universality among these unpolarized threshold TMDs (TTMDs). Explicitly, we find
\begin{align}
    f_{q/p}^{\rm TTMD}(x, \bm k_T, Q)\big |_{\rm SIDIS} &= f_{q/p}^{\rm TTMD}(x, \bm k_T, Q)\big |_{\rm DY}, \\
    D_{h/q}^{\rm TTMD}(z, \bm p_T, Q)\big |_{\rm SIDIS} &= D_{h/q}^{\rm TTMD}(z, \bm p_T, Q)\big |_{e^+e^-}.
\end{align}
This property also appears in the standard TMD factorization and resummation formula, which is very useful in the global fitting of different sets of TMD functions. In principle, such property of universality can also be generalized to the polarized TMD functions, which will be discussed thoroughly in future work. In this paper, we present the numerical results on the unpolarized threshold-TMD functions and also the transverse momentum cross section in three processes. In numerics, we restrict ourselves to resummation at NLL, which captures the main effects in the QCD evolution.  

The rest of this paper is organized as follows. In section \ref{sec:factorization}, we first review the factorization theorem in the joint threshold and small transverse momentum limit for the Drell-Yan process. Then, we introduce the definition of threshold-TMD PDFs and write down the corresponding QCD evolution function. In this section, we also briefly show the factorization theorem for SIDIS and $e^+e^-$ processes and give the definition of threshold-TMD FFs.  We present the numerical results in section \ref{sec:numerics}, where we first give the numerical predictions of the transverse momentum distribution for threshold-TMD PDFs and FFs, and then discuss the cross section for Drell-Yan, SIDIS and also $e^+e^-$. We conclude in section \ref{sec:conclusion}. The details of perturbative results of the collinear-soft function are provided in the appendix \ref{app:csfun}. We also collect the fitting parameters of PDFs and FFs used in our numerics in the appendix \ref{app:PDF_FF}. 

\section{Factorization and resummation formalism}\label{sec:factorization}
This section presents the factorization theorems for processes, characterized by the leading order partonic reaction $q \bar q \rightarrow \gamma^*$ or one of its crossed versions, in the joint threshold and small transverse momentum limits. First, we show the factorization theorem for the Drell-Yan process and present our definition of threshold-TMD PDFs based on RG equations. Then, we generalize our results to SIDIS and $e^+e^-$ processes and introduce the definition of threshold-TMD FFs, which captures both soft gluons and TMD evolution effects for the inclusive hadron production.

\subsection{Theory formalism in Drell-Yan}
\label{sec:dy-th}

For simplicity, we consider the Drell-Yan process mediated by a virtual photon with time-like momentum $q^{\mu}$,
\begin{align}
 h_1(P_1) + h_2(P_2) \to \gamma^*(q) \to l^+ + l^- + X,
\end{align}
where $X$ denotes the undetected hadronic particles in the final state. The standard TMD factorization theorem reads
\begin{align}
    \frac{\mathrm{d}^4\sigma^{\rm DY}}{\mathrm{d}^2 \bm{q}_T \, \mathrm{d} Q^2 \mathrm{d}Y}  & =   \frac{\sigma_0^{\rm DY}}{s}  H^{\rm DY}(Q,\mu)  \\
    &\times \int \frac{\mathrm{d}^2 \bm{b}_T}{4\pi^2} e^{i\bm{q}_T \cdot \bm{b}_T} \sum_{q} e_q^2 f_{q/h_1}^{\,\rm TMD}(x_1,b_T,\mu,\zeta) f_{\bar{q}/h_2}^{\,\rm TMD}(x_2,b_T,\mu,\zeta) + \mathcal{O}(\bm q_T^2/Q^2), \notag
\end{align}
where $\bm q_T$, $Q$ and $Y$ denote the transverse momentum, invariant mass and rapidity of final-state lepton pairs respectively. The differential cross section is factorized in terms of hard factor $H(Q,\mu)$ and TMD PDFs $f^{\rm TMD}_{q/h_i}(x_i,b_T,\mu,\zeta)$ from two colliding beams where $\mu$ and $\zeta$ denote the factorization and Collins-Soper scale respectively, and the soft function has been subtracted as in the standard redefinition procedure \cite{Collins:2011zzd,Ebert:2019okf}. Here $q$ runs over quarks and anti-quarks participating in the hard scattering, and $e_q$ denotes the corresponding charges. The light-cone momentum fractions of the hadron $h_{1,2}$ carried by the partons are denoted by $x_{1,2}$, which can be expressed as
\begin{align}\label{eq:x1x2}
    x_1 = \sqrt{\tau^{\rm DY} } e^{-Y},~~~x_2 = \sqrt{\tau^{\rm DY}} e^Y,~~~\mathrm{with}~~~\tau^{\rm DY}\equiv\frac{Q^2}{s}~~~\mathrm{and}~~~ s\equiv(P_1+P_2)^2.
\end{align}
In addition, for simplicity, we define the Born cross section as
\begin{align}
    \sigma_0^{\rm DY} \equiv \frac{4\pi \alpha_{\rm em}^2}{3N_c Q^2},
\end{align}
with the number of color $N_c=3$ and the fine structure constant $\alpha_{\rm em}$, so that the leading-order hard function $H^{\rm DY}(Q,\mu)$ is normalized to unity. The perturbative calculation of TMD PDFs using the operator-product expansion method shows that large logarithmic terms will appear in the limit $x \to 1$ \cite{Catani:2011kr,Catani:2012qa,Gehrmann:2012ze,Gehrmann:2014yya,Luebbert:2016itl,Echevarria:2016scs,Luo:2019hmp,Luo:2019bmw,Behring:2019quf,Luo:2019szz,Luo:2020epw,Ebert:2020yqt}. Therefore, one needs to consider the factorization of the cross section in the joint threshold and TMD limit.

Following~\cite{Sterman:2006hu} and performing the Mellin transformation with respect to the threshold variable $\tau^{\rm DY}$, we obtain
\begin{align}
    \frac{\mathrm{d}^2\tilde\sigma(N)}{\mathrm{d}^2 \bm q_T} &= \int_0^1 \mathrm{d} \tau^{\rm DY} \left(\tau^{\rm DY}\right)^{N-1} \int \mathrm{d} Y \mathrm{d} Q^2 \frac{\mathrm{d}^4\sigma}{\mathrm{d}^2 \bm{q}_T \, \mathrm{d} Q^2 \mathrm{d}Y} \delta(\tau^{\rm DY} - x_1 x_2) , \notag \\
    &= s\int d x_1 d x_2 (x_1 x_2)^{ N-1} \frac{\mathrm{d}^4\sigma}{\mathrm{d}^2 \bm{q}_T \, \mathrm{d} Q^2 \mathrm{d} Y},
\end{align}
where from the first line to the second line the Jacobian for converting between $(Q^2,Y)$ and $(x_1,x_2)$ can be easily worked out using the definitions of the kinematic variables in \eqref{eq:x1x2}. Therefore, the differential cross section can be re-expressed as
\begin{align}\label{eq:TMD_DY}
     \frac{\mathrm{d}^3\sigma^{\rm DY}}{\mathrm{d}^2 \bm{q}_T \, \mathrm{d} \tau^{\rm DY}}  =   \sigma_0^{\rm DY} \int_{C_N}\frac{\mathrm{d} N}{2\pi i}  \left(\tau^{\rm DY}\right)^{-N} &  \int \frac{\mathrm{d}^2 \bm{b}_T}{4\pi^2} e^{i\bm{q}_T \cdot \bm{b}_T} \, H^{\rm DY}(Q,\mu)  \\
     &\times \sum_{q} e_q^2\tilde{f}_{q/h_1}^{\,\rm TMD}(N,b_T,\mu,\zeta) \tilde{f}_{\bar{q}/h_2}^{\,\rm TMD}(N,b_T,\mu,\zeta), \notag
\end{align}
where we have applied the inverse Mellin transformation $\int_{C_N}\cdots$ to obtain the cross section in the momentum space. In the moment space, TMD PDF is defined as
\begin{align}
    \tilde{f}_{i/h}^{\,\rm TMD}(N,b_T,\mu,\zeta) \equiv \int_0^1 \mathrm{d} x \, x^{N-1} f_{i/h}^{\, \rm TMD}(x,b_T,\mu,\zeta).
\end{align}
 It is worth emphasizing that such transformation is not necessary for the standard TMD factorization theorem, but we keep it for the following joint threshold and TMD factorization. 

Now we briefly discuss the factorization formula in the joint threshold and TMD limit in SCET formalism. More detailed discussions on the NLL resummation formula can be found in \cite{Kulesza:2002rh,Kulesza:2003wn}, and the factorization formula within SCET is given in \cite{Lustermans:2016nvk,Li:2016axz}.  The threshold effects embody the behavior of the cross sections at $\tau \to 1$, since near the machine threshold $s \approx Q^2$, the colliding energy is just sufficient enough to produce the final-state lepton pair with the invariant mass $Q$. Therefore, one immediately realizes that the threshold limit in momentum space $\tau \to 1$ corresponds to $N \to \infty$ in Mellin space. This means the above factorization is incomplete in the sense that the threshold kinematics are ignored. Generally, there is a standard operator expansion step to relate TMD PDFs to collinear PDFs at small $b_T$~\cite{Collins:1981uw,Collins:1984kg,Collins:2011zzd}. Explictly, we have
\begin{align}
    f^{\rm TMD}_{i/h}\left(x, b_T, \mu, \zeta\right)=\sum_j \int_x^1 \frac{\mathrm{d} y}{y} C_{i j}\left(z, b_T, \mu, \zeta\right) f_{j/h}(y/z, \mu)+\mathcal{O}(b_T^2 \Lambda_{\rm QCD}^2),
\end{align}
where $C_{ij}(z,b_T,\mu,\zeta)$ are perturbative matching coefficients. It is known that to arbitrary order there are threshold logarithms $[\ln^n(1-z)/(1-z)]_{+}$ in $C_{ij}$ as $z\to 1$, which must be resummed to achieve precise results. To incorporate both the TMD and threshold effects, we employ the re-factorization technique presented in \cite{Li:2016axz,Lustermans:2016nvk} for the TMD PDF $\tilde{f}_{i/h}^{\,\rm TMD}$ 
\begin{align}\label{eq:TMDPDF_largeN}
    \tilde{f}_{i/h}^{\,\rm TMD} (N,b_T,\mu,\zeta) \xrightarrow[]{N\to \infty} \, \underbrace{\tilde{S}_c^{\rm unsub}(b_T,\mu,\zeta_N/\nu^2) \sqrt{S(b_T,\mu,\nu)}}_{\let\scriptstyle\textstyle\substack{\equiv \, \tilde{S}_{c}(b_T,\mu,\zeta_N)}}\tilde{f}_{i/h}(N,\mu) + \mathcal{O}(b_T^2 \Lambda_{\rm QCD}^2), 
\end{align}
where we introduce the Collins-Soper scale $\zeta_N\equiv\zeta/\bar N^2$ with $\bar N = N e^{\gamma_E}$ and the Euler constant $\gamma_E$ in the threshold limit. The rapidity scale dependence $\nu$ is cancelled between the Mellin space unsubtracted collinear-soft function $\tilde{S}_c^{\rm unsub} (b_T, \mu, \zeta_N / \nu^2) $ and the standard TMD soft function $S(b_T,\mu,\nu)$. We refer to $\tilde{S}_{c}(b_T,\mu,\zeta_N)$ as the genuine collinear-soft function in the joint threshold and TMD limit, which is flavor and spin-independent, but different for quarks and gluons. In the appendix \ref{app:csfun}, we provide the derivation of its one-loop expression in the perturbative region and also discuss its relation to the perturbative matching coefficient of TMD PDFs in threshold limit. Moreover, we derive its expressions up to the three-loop order based on the threshold behaviors of the next-to-next-to-next-to-leading order (NNNLO) perturbative matching coefficients of TMD PDFs \cite{Luo:2019szz,Luo:2020epw,Ebert:2020yqt}. In Eq.~\eqref{eq:TMDPDF_largeN}, $\tilde{f}_{i/h}(N,\mu)$ is the collinear PDF in Mellin space, which takes the form
\begin{align}
    \tilde{f}_{i/h}(N,\mu) = \int_0^1 \mathrm{d} x \, x^{N-1} f_{i/h}(x,\mu). 
\end{align}
It is important to emphasize that the aforementioned refactorization formula \eqref{eq:TMDPDF_largeN} is only proven for small values of $b_T$ \cite{Li:2016axz,Lustermans:2016nvk}. The generalization of this formula to higher-twist distributions remains unclear, and further investigation is required to address this issue. We leave the study of higher-twist distributions for future works.

Since both the unsubtracted collinear-soft and TMD soft function depend on the rapidity scale $\nu$, their rapidity scale evolution is governed by the following $\nu$-RG evolution equations~\cite{Chiu:2011qc,Chiu:2012ir}
\begin{align}
    \frac{\mathrm{d}}{\mathrm{d} \ln{\nu}}\tilde{S}_c^{\rm unsub}(b_T,\mu,\zeta_N/\nu^2) &= \gamma^{S_c}_\nu(b_T,\mu) \tilde{S}_c^{\rm unsub}(b_T,\mu,\zeta_N/\nu^2)\,, \label{eq:cs-rrg}\\
    \frac{\mathrm{d}}{\mathrm{d} \ln{\nu}}S(b_T,\mu,\nu) &= \gamma^{S}_\nu(b_T,\mu) S(b_T,\mu,\nu)\,, \label{eq:cs-rrg0}
\end{align}
with $\gamma^{S_c}_\nu=-\gamma^{S}_\nu/2$. Upon subtracting the soft function, as previously outlined in Eq.~\eqref{eq:TMDPDF_largeN}, we have the genuine collinear-soft function $\tilde{S}_c(b_T,\mu,\zeta_N)$. In particular, we find that $\tilde{S}_c^{\rm unsub}(b_T,\mu,\zeta_N/\nu^2)$ depends only on the combination $\zeta_N/\nu^2$ in the threshold limit~\footnote{Note that away from the threshold limit, the unsubstracted TMD PDFs usually depend on the combination $\zeta/\nu^2$~\cite{Ebert:2019okf,Boussarie:2023izj}.}, which is verified up to the three-loop order in appendix \ref{app:csfun}. This allows us to convert the above two $\nu$-RG evolution equations in Eqs.~\eqref{eq:cs-rrg} and \eqref{eq:cs-rrg0} into a similar Collins-Soper evolution equation for genuine collinear-soft function $\tilde{S}_{c}(b_T,\mu,\zeta_N)$ by following the procedure outlined in~\cite{Ebert:2019okf}
\begin{equation}\label{eq:cseqn}
    \sqrt{\zeta_N} \frac{\mathrm{d}}{\mathrm{d} \sqrt{\zeta_N} }\,\tilde{S}_{c}(b_T,\mu,\zeta_N) =\kappa ( b_T,\mu )\,\tilde{S}_{c} (b_T,\mu,\zeta_N),
\end{equation}
where $\kappa ( b_T,\mu ) = -\gamma_\nu^{S_c}(b_T,\mu)$ is the Collins-Soper kernel~\cite{Collins:2011zzd,Ebert:2019okf} or the rapidity anomalous dimension~\cite{Chiu:2011qc,Chiu:2012ir} and collinear anomaly exponent~\cite{Becher:2010tm,Becher:2011xn} in SCET. Here, from perturbative calculation, $\kappa ( b_T,\mu )$ can be written as $\kappa ( b_T,\mu ) =  -\Gamma _{\text{cusp}}( \alpha _{s}) L_{b} +\mathcal{O}\left( \alpha _{s}^{2}\right)$, where $L_b=\ln (\mu^2 b_T^2/b_0^2)$ with $b_0=2e^{-\gamma_E}$. We have verified Eq.~\eqref{eq:cseqn} up to three-loop order based on the NNNLO expression of $\tilde{S}_c$ given in Eq. \eqref{eq:cs-threeloop}. The four-loop perturbative expression of $\kappa(b_T,\mu)$ was recently calculated in \cite{Moult:2022xzt,Duhr:2022yyp}, and its nonperturbative analysis can be found in \cite{Becher:2013iya,Vladimirov:2020umg} and the preliminary nonperturbative numerical simulation in lattice QCD is presented in \cite{Shanahan:2021tst,LPC:2022ibr,LatticePartonLPC:2023pdv}.

It is noteworthy that the Collins-Soper equation governing $\tilde{S}_{c}$ as a function of $\sqrt{\zeta_N}$ coincides with the equation for TMD PDFs as a function of $\sqrt{\zeta}$. This correspondence arises from the ($\nu$-)RG invariance analysis elucidated in Ref.~\cite{Lustermans:2016nvk}. Specifically, the rapidity divergence remains the same in the threshold limit, defined by $Q \gg Q(1- \hat{\tau}) \gg q_T$, where $\hat{\tau}$ is the partonic analog of $\tau^{\text{DY}}$. Consequently, the rapidity-RG equation \eqref{eq:cs-rrg} for the unsubtracted collinear-soft function is unchanged. It is crucial, however, to highlight that the rapidity scale differs between the unsubtracted collinear-soft function and the unsubtracted TMD PDFs. In essence, all threshold logarithms present in the matching coefficients of TMD PDFs are subsumed into the rapidity logarithms. This has been rigorously confirmed up to three-loop calculations~\cite{Luo:2019szz,Luo:2020epw,Ebert:2020yqt}. Therefore, we introduce a modified Collins-Soper scale $\zeta_N$ in the joint threshold and TMD limit.

After solving the above Collins-Soper evolution equation for the $\zeta_N$ dependence at the renormalization scale $\mu = \mu_b$, we have 
\begin{align}\label{eq:cs-evolve}
    \tilde{S}_{c}\left(b_T, \mu_b, \zeta_{N,f}\right)=\tilde{S}_{c}\left(b_T, \mu_b, \zeta_{N,i}\right) \left(\sqrt{\frac{\zeta_{N,f}}{\zeta_{N,i}}}\right)^{\kappa\left(b_T,\, \mu_{b}\right)},
\end{align}
where we can choose $ \zeta_{N,i} =\mu _{b}^{2} =b_{0}^{2} /b_T^{2}$ and $ \zeta _{N,f}$ will be determined from RG consistency \cite{Collins:2011zzd,Ebert:2019okf}. To obtain its value, we first write down RG equations for the hard function, collinear-soft function and threshold PDFs
\begin{align}
    \mu \frac{\mathrm{d}}{\mathrm{d} \mu } H^{\rm DY}\left( Q ,\mu \right) & = \Gamma^h(\alpha_s) H^{\rm DY}\left( Q ,\mu \right), \\
    \mu \frac{\mathrm{d}}{\mathrm{d} \mu } \tilde{S}_c\left( b_T,\mu,\zeta_{N,f} \right) & = \Gamma^{\tilde{S}_c}(\alpha_s,\zeta_{N,f}) \tilde{S}_c\left( b_T ,\mu,\zeta_{N,f} \right), \\
    \mu \frac{\mathrm{d}}{\mathrm{d} \mu } \tilde{f}_q\left( N,\mu \right) & = \Gamma^{\tilde{f}_q}(\alpha_s) \tilde{f}_q\left( N ,\mu \right),
\end{align} 
and the corresponding anomalous dimensions are
\begin{align}
    \Gamma^h(\alpha_s) & =  2\,\Gamma _{\text{cusp}}( \alpha _{s})\,\mathrm{ln}\frac{Q^{2}}{\mu ^{2}} +2\gamma _{V}( \alpha _{s})\,, 
    \\
    \Gamma^{\tilde{S}_c}(\alpha_s,\zeta_{N,f}) & =  -\Gamma _{\text{cusp}}( \alpha _{s})\,\mathrm{ln}\frac{\zeta _{N,f}}{\mu ^{2}} +\gamma _{\tilde{S}_c}( \alpha _{s})\,, 
    \\
    \Gamma^{\tilde{f}_q}(\alpha_s) & = -2\,\Gamma _{\text{cusp}}( \alpha _{s})\,\mathrm{ln}\bar{N} +2\gamma _{\tilde{f}_q}(\alpha_s)\,,
\end{align}
with perturbative coefficients of anomalous dimensions needed at the NLL accuracy\footnote{Here, we provide the two-loop cusp anomalous dimensions that are employed in $\Gamma^h$ and $\Gamma^{\tilde f_i}$. However, it is important to note that, for the Collins-Soper kernel $\kappa$, we consistently utilize its one-loop results in our NLL resummation calculations.} as
\begin{align}\label{eq:singlelog}
    &\gamma ^{V}_{0} =-6\,C_{F} ,~~~ \gamma ^{\tilde{f}_q}_{0} =3\,C_{F},~~~   \gamma ^{\tilde{S}_c}_{0} =0, \\
    & \Gamma_0 = 4\, C_F, ~~~ \Gamma_1= \left(\frac{268}{9}-\frac{4 \pi^{2}}{3}\right) C_F C_{A}-\frac{40}{9} C_{F} n_{f},
\end{align}    
where they are defined by
\begin{align}
    \Gamma_{\rm cusp}(\alpha_s) = \sum_{n=0}^\infty \Gamma_{n} \left(\frac{\alpha_s}{4\pi}\right)^{n+1}\,,
\qquad
\gamma_{V, \, \tilde{S}_c, \,\tilde{f}_q}(\alpha_s) = \sum_{n=0}^\infty \gamma_{n}^{V, \, \tilde{S}_c,\,\tilde{f}_q} \left(\frac{\alpha_s}{4\pi}\right)^{n+1}\,.
\end{align}
Then the RG consistency $\Gamma^h +2 \, \Gamma^{\tilde{S}_c} + 2 \, \Gamma^{\tilde{f}_q}=0$ implies that 
\begin{align}\label{eq:zetaf}
    \zeta_{N,f} =\frac{Q^{2}}{\bar{N}^{2}}.
\end{align}
We observe that the value of $\zeta _{N,f}$ is different from the one in the standard TMD factorization formula. In the momentum space the scale $\zeta _{N,f}$ can be expressed as $\zeta_{N,f}\sim Q^2(1-\hat \tau)^2$. We note that its value is reduced from the original Collins-Soper scale $\zeta_{f}= Q^2$ in the standard TMD factorization. This is expected since we have taken into account the threshold effects in the joint factorization formula, and the phase space for the initial collinear radiations is further constrained in the threshold limit. Especially, as $Q(1-\hat \tau) \to q_T$, i.e. $\zeta_{N,f} \to \zeta_{N,i}$, the rapidity evolution effects in \eqref{eq:cs-evolve} will be turned off automatically.

All-order resummation formula can be obtained by solving the RG equations in both position and moment spaces and evolving the ingredients from their intrinsic scales to a common scale. The all-order resumed cross section is given as
\begin{align}
     \frac{\mathrm{d}^3\sigma^{\rm DY}}{\mathrm{d}^2\bm{q}_T \, \mathrm{d}\tau^{\rm DY}}  =   \sigma_0^{\rm DY} \int_{C_N}\frac{\mathrm{d} N}{2\pi i}  \left(\tau^{\rm DY}\right)^{-N} &
     \int_0^\infty \frac{\mathrm{d} b_T \, b_T}{2 \pi} J_0(q_T \, b_T) H^{\rm DY}(Q,Q) \\
     & \times \sum_{q} e_q^2 \tilde{f}_{q/h_1}^{\,\rm TTMD}(N,b_T,Q) \tilde{f}^{\,\rm TTMD}_{\bar q/h_2}(N,b_T,Q), \notag
\end{align}
where we introduce a new type of TMD PDFs, i.e. the \textit{threshold-TMD PDFs} $\tilde{f}^{\,\rm TTMD}_{i/h}$, and then the cross section could be factorized as the product of the hard function and threshold-TMD PDFs. The definition of the threshold-TMD PDF at a scale of $Q$ is
\begin{align}\label{eq:thrshold_TMDPDF}
    \tilde{f}^{\,\rm TTMD}_{i/h}(N,b_T,Q) = & \exp\left[ -S_{\rm pert}(Q,\mu_{b_*},\mu_F)-S_{\rm NP}^{f} \left(b_T,Q_0,\zeta_{N,f}\right) \right] \tilde f_{i/h}(N,\mu_F),
\end{align}
where the perturbative evolution kernel $S_{\rm pert}$ involves the contribution from both collinear PDF and collinear-soft functions, which reads 
\begin{align}\label{eq:TTMD_Spert}
    S_{\rm pert}(Q,\mu_{b_*},\mu_F) = \int_{\mu_{b_*}}^{Q} \frac{d\mu}{\mu} \frac{\Gamma^h(\alpha_s)}{2}-\int_{\mu_F}^{\mu_{b_*}}\frac{d\mu}{\mu}\Gamma^{\tilde f_i}(\alpha_s) - \frac{1}{2}\kappa(b_*,\mu_{b_*})\ln \frac{\zeta_{N,f}}{\zeta_{N,i}}.
\end{align}
Here note that the $\Gamma^h$ term is the same as the standard TMD PDFs since the hard function only receives virtual pQCD corrections and is unchanged in the threshold limit. The $\Gamma^{\tilde f_i}$ term may be viewed as a simplified DGLAP equation, where the flavor off-diagonal pieces are non-singular as $N\to \infty$, and thus can be neglected at the leading power formalism. Besides, $\mu_{b_*}$ and $\mu_F$ are the intrinsic scales of collinear-soft function and collinear PDFs. We stress that the above perturbative evolution kernel is consistent with that in NLL resummation formula \cite{Kulesza:2002rh,Kulesza:2003wn}, and it can also be evaluated beyond the NLL accuracy \cite{Lustermans:2016nvk} after including higher order ingredients within the perturbative QCD framework. 

To match the perturbative and non-perturbative contributions in \eqref{eq:thrshold_TMDPDF}, we first apply the standard $b_*$-prescription \cite{Collins:1984kg} to avoid the Landau pole in the non-perturbative region as $b_T\to \infty$, which is
\begin{align}
    b_* \equiv \frac{b_T}{\sqrt{1+b_T^2/b_{\rm max}^2}},~~~~ {\rm with}~~ \mu_{b_*}=b_0/b_*~~{\rm and}~~b_{\rm max} = 1.5\, {\rm GeV}^{-1}.
\end{align}
In addition to the $b_*$-prescription, we also apply the model in \cite{Su:2014wpa,Echevarria:2020hpy} to parametrize the non-perturbative contribution at large $b_T$. Explicitly, in \eqref{eq:thrshold_TMDPDF}, we define the non-perturbative kernel as
\begin{align}\label{eq:SNP_DY}
    S_{\mathrm{NP}}^f \left(b_T, Q_{0}, \zeta_{N,f} \right)=g_{1}^f b_T^{2}+\frac{g_{2}}{2}  \ln \frac{\sqrt{\zeta_{N,f}}}{Q_{0}} \ln \frac{b_T}{b_{*}}
\end{align}
with $g_1^f=0.106 \, {\rm GeV}^2$, $g_2=0.84$ and $Q_0^2=2.4 \,{\rm GeV}^2$.  We observe that the value of $ \zeta_{N,f}$   in the threshold limit deviates from its counterpart in standard TMD resummation, as elucidated subsequent to Eq.~\eqref{eq:zetaf}. Accordingly, in Eq.~\eqref{eq:SNP_DY}, we opt for the Collins-Soper scale factor $  Q/\bar{N} $  rather than $Q$ as originally posited in Ref.~\cite{Su:2014wpa}. We must underscore the fact that our adaptation of the model does not account for the complete non-perturbative threshold enhancement effects. Therefore, additional fits in the threshold region are requisite for achieving high-precision results. Lastly, apart from the non-perturbative factor $ S_{\rm NP} $ associated with TMDs, it is necessary to incorporate the collinear PDF, encapsulated in the factor $ \tilde{f}_{i/h}(N,\mu_F) $, defined at the scale $ \mu_F $.

\subsection{Theory formalism in SIDIS and $e^+e^-$}
\label{sec:dia-th}
In addition to TMD PDFs, another important set of distributions for probing hadronic three-dimensional structures is the TMD FFs, which can be studied in SIDIS and back-to-back two hadron production in $e^+ e^-$ collisions, separately. Similar to the structure of the previous subsection, we first review the factorization in terms of usual threshold variables in SIDIS and $e^+ e^-$. Then we define the TMD FFs in the threshold limit. As we will see, our formalism has an advantage over the usual pure threshold formalism allowing further discussions. Finally, we present the factorization for SIDIS and $e^+ e^-$ in the joint threshold and TMD limit and a collected version.

For the SIDIS process, we consider a proton $p$ with momentum $P_1^\mu$ is probed by a virtual photon with space-like momentum $q^\mu$ and produces a final-state inclusive hadron $h$ with momentum $P_2^\mu$. Explicitly, we have
\begin{equation}
e^-(\ell) +p( P_{1})\rightarrow e^-(\ell ^\prime) +h( P_{2}) +X,
\end{equation}
which probes the short-distance scattering of the electron and a quark inside the proton $p$ by exchanging a virtual photon. The standard unpolarized differential cross section is~\cite{Bacchetta:2006tn,Collins:2011zzd,Echevarria:2020hpy}
\begin{align}
    \frac{\mathrm{d} \sigma^{\rm SIDIS}}{\mathrm{d} x \mathrm{d} y \mathrm{d} z \mathrm{d}^2 \boldsymbol{q}_T } & = \sigma_0^{\rm DIS}  H^{\rm SIDIS}(Q, \mu)   \\ & \times \int \frac{\mathrm{d} ^2 \boldsymbol{b}_T}{4 \pi^2} e^{i \boldsymbol{q}_T \cdot \boldsymbol{b}_T}  \sum_{q} e_{q}^{2} f_{q/p}^{\rm TMD} (x,b_T, \mu, \zeta) D_{h/q}^{\rm TMD}(z, b_T, \mu, \zeta), \notag 
\end{align}
where $ \boldsymbol{q}_T $ is the transverse momentum of the photon in the hardon proton frame, and $\sigma_0^{\rm DIS}$ is the leading order electromagnetic scattering cross section given by
\begin{equation}
    \sigma_0^{\rm DIS} = \frac{2 \pi \alpha_{\rm em}^2}{Q^2} \frac{1+(1-y)^2}{y}.
\end{equation}
Besides, $H^{\mathrm{SIDIS}}$ denotes the hard function in the SIDIS process, and $D^{\rm TMD}_{h/q}$ is the standard TMD FF. Note that we have included a factor of $z^2$ into the definition of $D^{\rm TMD}_{h/q}(z, b_T, \mu, \zeta)$. Here, we have employed the usual SIDIS kinematic variables
\begin{equation}
    x = \frac{Q^2}{2P_1 \cdot q}, \ \  y= \frac{Q^2}{xs},  \ \ z= \frac{P_1 \cdot P_2}{P_1 \cdot q},
\end{equation}
with $q = \ell - \ell^{\prime} , \ \ Q^2 = - q^2 $ and $s=(P_1+\ell)^2.$ In the Mellin space, the above factorization takes the form
\begin{align}
    \frac{\mathrm{d}^3\tilde\sigma(N)}{\mathrm{d} y \mathrm{d}^2 \bm q_T} &= \int_0^1 \mathrm{d} \tau^{\rm SIDIS} \left(\tau^{\rm SIDIS}\right)^{N-1} \int \mathrm{d}x \mathrm{d}z \, \delta\left(\tau^{\rm SIDIS} - x z\right) \frac{\mathrm{d}^5\sigma}{\mathrm{d}x\mathrm{d}y \mathrm{d}z\mathrm{d}^2 \bm{q}_T }, \notag \\
    &= \int \mathrm{d} x \mathrm{d} z (x z)^{ N-1} \frac{\mathrm{d}^5\sigma}{\mathrm{d}x \mathrm{d} y \mathrm{d} z \mathrm{d}^2 \bm{q}_T   },
\end{align}
where, following \cite{Sterman:2006hu}, we have $\tau^{\rm SIDIS} \equiv xz$ that is very similar to the Drell-Yan threshold variable $\tau^{\rm DY}$. For this reason, this is referred to as ``crossed threshold variable'' in~\cite{Sterman:2006hu}. After the inverse Mellin transformation, the above factorization can be rewritten as 
\begin{align}
    \frac{\mathrm{d}^4 \sigma^{\rm SIDIS}}{\mathrm{d} y \mathrm{d}^2 \boldsymbol{q}_T d\tau^{\rm SIDIS}} = \sigma_0^{\rm DIS} & \int_{C_N} \frac{\mathrm{d}N}{2\pi i}\big(\tau^{\rm SIDIS}\big)^{-N} \int \frac{\mathrm{d} ^2 \boldsymbol{b}_T}{4 \pi^2} e^{i \boldsymbol{q}_T \cdot \boldsymbol{b}_T} H^{\rm SIDIS}(Q, \mu) \\ & \times \sum_{q} e_{q}^{2} \tilde{f}_{q/p}^{\,\rm TMD} (N,b_T, \mu, \zeta) \tilde{D}_{h/q}^{\rm TMD}(N, b_T, \mu, \zeta). \notag 
\end{align}
In the threshold limit i.e $\tau^{\rm SIDIS} \to 1$ in momentum space or $N \to \infty$ in Mellin space, 
the above factorization is incomplete. One needs to re-factorize the TMD FFs to take into account this threshold effect. We find that the re-factorized formula can be written in Mellin space as 
\begin{equation}
    \tilde{D}_{h/q}^{\rm TMD} (N,b_T,\mu, \zeta) \xrightarrow[]{N\to \infty} \, \underbrace{\tilde{S}_c^{\rm{unsub }}\left(b_T, \mu, \zeta_N / \nu^2\right) \sqrt{S(b_T, \mu, \nu)}}_{\let\scriptstyle\textstyle\substack{\equiv \, \tilde{S}_c(b_T, \mu, \zeta_N)}} \tilde{D}_{h / q}(N, \mu)+\mathcal{O}(b_T^2\Lambda_{\rm QCD}^2),
\end{equation}
where $ \tilde{D}_{h/q} (N, \mu) $ is the collinear FF in Mellin space. It is not surprising to find that the genuine collinear-soft functions for TMD FFs are the same as the ones in TMD PDFs. Since, in the threshold limit, the longitudinal momentum fractions of TMD FFs $z \to 1$, making the behavior of the TMD FFs similar to that of the TMD PDFs at NLL. Moreover, we stress that it is one of the advantages of our joint threshold and TMD formalism to allow one to define threshold distributions separately.

To obtain the Collins-Soper scale, one exploits the RG consistency by
\begin{align}
    \mu \frac{\mathrm{d}}{\mathrm{d} \mu } H^{\rm SIDIS}\left( Q ,\mu \right) & = \Gamma^h(\alpha_s) H^{\rm SIDIS} \left( Q ,\mu \right), \\
    \mu \frac{\mathrm{d}}{\mathrm{d} \mu } \tilde{S}_c\left( b_T,\mu,\zeta_{N,f} \right) & = \Gamma^{\tilde{S}_c}(\alpha_s,\zeta_{N,f}) \tilde{S}_c\left( b_T ,\mu,\zeta_{N,f} \right), \\
    \mu \frac{\mathrm{d}}{\mathrm{d} \mu } \tilde{D}_q\left( N,\mu \right) & = \Gamma^{\tilde{D}_q}(\alpha_s) \tilde{D}_q\left( N ,\mu \right), \\
    \mu \frac{\mathrm{d}}{\mathrm{d} \mu } \tilde{f}_q\left( N,\mu \right) & = \Gamma^{\tilde{f}_q}(\alpha_s) \tilde{f}_q\left( N ,\mu \right)
\end{align} 
and the corresponding anomalous dimensions are
\begin{align}
    \Gamma^h(\alpha_s) & =  2\,\Gamma _{\text{cusp}}( \alpha _{s})\,\mathrm{ln}\frac{Q^{2}}{\mu ^{2}} +2\gamma _{V}( \alpha _{s}) \\
    \Gamma^{\tilde{S}_c}(\alpha_s,\zeta_{N,f}) & =  -\Gamma _{\text{cusp}}( \alpha _{s})\,\mathrm{ln}\frac{\zeta_{N,f}}{\mu ^{2}} +\gamma _{\tilde{S}_c}( \alpha _{s}) \\
    \Gamma^{\tilde{D}_q}(\alpha_s) & = -2\,\Gamma _{\text{cusp}}( \alpha _{s})\,\mathrm{ln}\bar{N} +2\gamma _{\tilde{D}_q}( \alpha _{s}) \\
    \Gamma^{\tilde{f}_q}(\alpha_s) & = -2\,\Gamma _{\text{cusp}}( \alpha _{s})\,\mathrm{ln}\bar{N} +2\gamma _{\tilde{f}_q}( \alpha _{s})
\end{align}
with perturbative coefficients of anomalous dimensions needed at the NLL accuracy as
\begin{align}\label{eq:singlelog}
    &\gamma ^{V}_{0} =-6\,C_{F} ,~~~ \gamma ^{\tilde{D}_q}_{0} =3\,C_{F} = \gamma ^{\tilde{f}_q}_{0},~~~   \gamma ^{\tilde{S}_c}_{0} =0, \\
    & \Gamma_0 = 4\, C_F, ~~~ \Gamma_1= \left(\frac{268}{9}-\frac{4 \pi^{2}}{3}\right) C_F C_{A}-\frac{40}{9} C_{F} n_{f}.
\end{align}
Then $ \Gamma^h + (\Gamma^{\tilde{S}_c} + \Gamma^{\tilde{D}_q}) + (\Gamma^{\tilde{S}_c} + \Gamma^{\tilde{f}_q}) = 0 $ just implies exactly the same Collins-Soper scale as that in the Drell-Yan process

\begin{equation}
     \zeta_{N,f} = \frac{Q^2}{\bar{N}^2}.
\end{equation}
Therefore, we define the threshold-TMD FFs in the Mellin moment space at a scale $Q$ as
\begin{equation}\label{eq:TTMDFF}
    \tilde{D}_{h/i}^{\mathrm{TTMD}}\left(N, b_T, Q\right)=\exp \left[-S_{\text {pert }}\left(Q, \mu_{b_*}, \mu_F\right)-S_{\mathrm{NP}}^D\left(b_T, Q_0, \zeta_{N,f}\right)\right] \tilde{D}_{h / i}\left(N, \mu_F\right),
\end{equation}
with the perturbative evolution kernel $ S_{\rm pert} $ is defined as
\begin{equation}
    S_{\rm pert} (Q,\mu_{b_*}, \mu_F)= \int _{\mu _{b_{*}}}^{Q}\frac{d\mu }{\mu }\frac{\Gamma ^{h}( \alpha _{s})}{2} -\int _{\mu_F}^{\mu _{b_{*}}}\frac{d\mu }{\mu } \Gamma ^{\tilde D_{i}}( \alpha _{s}) - \frac{1}{2}\kappa(b_*,\mu_{b_*})\ln \frac{\zeta_{N,f}}{\zeta_{N,i}}.
\end{equation}
The nonperturbative factor $S_{\rm NP}^D$ is defined as 
\begin{align}\label{eq:SNP}
    S_{\mathrm{NP}}^D \left(b_T, Q_{0}, \zeta_{N,f} \right)=g_{1}^D b_T^{2}+\frac{g_{2}}{2}  \ln \frac{\sqrt{\zeta_{N,f}}}{Q_{0}} \ln \frac{b_T}{b_{*}},
\end{align} 
with $g_1^D=0.042\, {\rm GeV}^2$ \cite{Su:2014wpa,Echevarria:2020hpy}. The values of $Q_0$ and $g_2$ have been given in \eqref{eq:SNP_DY}. It is noted that the above parametrization is consistent with the model used in \cite{Su:2014wpa,Echevarria:2020hpy} in the threshold limit $z \to 1$. Also, as mentioned earlier, a new fitting including the threshold effect is required for a higher precision theoretical result. Thus, the all-order resummation formula reads
\begin{align}\label{eq:resum_SIDIS}
\frac{\mathrm{d}^{4} \sigma ^{\mathrm{SIDIS}}}{\mathrm{d} y\mathrm{d}^{2}\boldsymbol{q}_{T}\mathrm{d} \tau ^{\mathrm{SIDIS}}} =\sigma _{0}^{\mathrm{DIS}}\int _{C_{N}}\frac{\mathrm{d} N}{2\pi i}\left( \tau ^{\mathrm{SIDIS}}\right)^{-N} & \int \frac{\mathrm{d}b_T \, b_T}{2\pi} J_0(q_{T} b_T) H^{\mathrm{SIDIS}}\left( Q ,Q \right)  \\
& \times \sum _{q} e_{q}^{2}\tilde{f}_{q/h_{1}}^{\,\mathrm{TTMD}} (N,b_T,Q )\tilde{D}_{h_{2} /q}^{\mathrm{TTMD}} (N,b_T,Q ). \notag
\end{align}

Finally, we move on to the factorization and resummation for the inclusive back-to-back two hadron production in $e^+e^-$ collision, i.e.
\begin{equation}
e^{+} +e^{-}\rightarrow \gamma ^{*}(q) \rightarrow h_{1}( P_{1}) +h_{2}( P_{2}) +X,
\end{equation}
where $P_i^\mu$ is the momentum of the hadron $h_i$, and the threshold variable in this process is defined by
\begin{equation}
    \tau^{e^+e^-} \equiv \frac{(P_1+P_2)^2}{Q^2},
\end{equation}
with $Q^2 = q^2$. As expected, in the joint limit, the cross section is factorized as the product of hard factor and two TMD FFs that describe final-state di-hadron production. Besides, the RG consistence also implies the Collins-Soper scale is the same as that in the Drell-Yan and SIDIS processes. Therefore, we obtain the following all-order resummation formula.
\begin{align}
    \frac{\mathrm{d}^{3} \sigma ^{e^+e^-}}{\mathrm{d}^{2}\boldsymbol{q}_{T}\mathrm{d} \tau ^{e^+e^-}} =\sigma _{0}^{e^+e^-}\int _{C_{N}}\frac{\mathrm{d} N}{2\pi i}\left( \tau ^{e^+e^-}\right)^{-N} & \int \frac{\mathrm{d}b_T \, b_T}{2\pi} J_0(q_{T} b_T) H^{e^+e^-}\left( Q ,Q \right) \\
    & \times \sum _{q} e_{q}^{2}\tilde{D}_{h_{1} /q}^{\mathrm{TTMD}} (N,b_T,Q)\tilde{D}_{h_{2} /\bar{q}}^{\mathrm{TTMD}} (N,b_T,Q ), \notag
\end{align}
where $\sigma _{0}^{e^+e^-} = 4 \pi \alpha_{\rm em}^2/Q^2$ is the Born cross section and $H^{e^+e^-}$ is the corresponding hard factor, and the definition of the threshold-TMD FFs at the scale $Q$ is given in \eqref{eq:TTMDFF}. 

From now on, we have obtained the all-order resummation formula for Drell-Yan, SIDIS and $e^+e^-$ processes, and also find a close correspondence between them. In the joint limit, the cross section is factorized as the product of the hard factor and threshold-TMD functions, including TMD PDFs and FFs. Such a property of universality is significant in the future global fitting analysis for threshold-TMD functions in different processes. To summarize this section, we give the following generic resummation structure for all three processes as
\begin{align}
    \frac{\mathrm{d}^{3} \sigma ^{\mathrm{A}}}{\mathrm{d}^{2}\boldsymbol{q}_{T}\mathrm{d} \tau ^{\mathrm{A}}} =\sigma _{0}^{\mathrm{A}}\int _{C_{N}}\frac{\mathrm{d} N}{2\pi i}\left( \tau ^{\mathrm{A}}\right)^{-N} & \int \frac{\mathrm{d}b_T \, b_T}{2\pi} J_0(q_{T} b_T) H^{\rm A}\left( Q ,Q \right) \\
    & \times \sum _{q} e_{q}^{2}\tilde{\mathcal{F}}_{i,h_1}^{\mathrm{TTMD}} (N,b_T,Q)\tilde{\mathcal{F}}_{i,h_2}^{\mathrm{TTMD}} (N,b_T,Q ), \notag
\end{align}
where for ${\rm A} = {\rm DY}$; $ \tilde{\mathcal{F}}_{i,h_1}^{\mathrm{TTMD}} = \tilde{f}_{q/h_{1}}^{\,\mathrm{TTMD}} $, $ \tilde{\mathcal{F}}_{i,h_2}^{\mathrm{TTMD}} = \tilde{f}_{\bar{q}/h_{2}}^{\,\mathrm{TTMD}} $, for $\rm A=SIDIS$; $ \tilde{\mathcal{F}}_{i,h_1}^{\mathrm{TTMD}} = \tilde{f}_{q/h_{1}}^{\,\mathrm{TTMD}} $, $ \tilde{\mathcal{F}}_{i,h_2}^{\mathrm{TTMD}} = \tilde{D}_{h_{2}/q}^{\,\mathrm{TTMD}} $, and for ${\rm A}=e^+e^-$; $ \tilde{\mathcal{F}}_{i,h_1}^{\mathrm{TTMD}} = \tilde{D}_{h_{1}/q}^{\,\mathrm{TTMD}} $, $ \tilde{\mathcal{F}}_{i,h_2}^{\mathrm{TTMD}} = \tilde{D}_{h_{2}/\bar{q}}^{\,\mathrm{TTMD}} $ and the Born cross sections $\sigma_0^{\rm A}$ and hard function $H^{\rm A}$ are defined accordingly. Here in addition to $\tau$, the Born cross section $\sigma_0$ can also depend on additional variables, for example, $\sigma_0^{\rm SIDIS}$ depends on the inelasticity $y$ as shown in \eqref{eq:resum_SIDIS}. This structure holds in the all-order QCD resummation formula.

\section{Numerical results}\label{sec:numerics}
In this section, we numerically study the threshold effect in TMD PDF and TMD FF using the factorization formula of Drell-Yan, SIDIS, and the back-to-back di-hadron production via the $e^+e^-$ annihilation process. We apply these extracted TMD functions to provide transverse momentum distributions for these three processes for different experimental kinematics. Throughout our numerical analysis, we use one loop strong coupling constant ($\alpha_s$). The fine structure constant ($\alpha_{\rm em}$) is taken to be $1/137$ and the number of active quark flavor $n_f = 5$ in the massless limit. For the TMD and collinear evolution, we choose our initial scale $(\mu_F)$ to be $1.3$ GeV. As mentioned earlier, threshold factorization has been done in the Mellin space because, in this space, the convolutions become a simple product. After achieving the factorization formula in Mellin space, we need to do the Mellin inversion to achieve results in the $x$ space. During our numerical study, we observe that one needs to modify the Collins-Soper scale in order to achieve results that are consistence throughout the allowed kinematic region of $\tau$. This modified Collins-Soper scale will introduce new poles in the Mellin inversion formula. We discuss these issues in detail in the next subsection and present a prescription for Mellin inversion. 

\subsection{Modified Collins-Soper scale and inverse Mellin transformation}

In the NLL resummation formula, the Collins-Soper scale in the non-perturbative factor $S_{\rm NP}$ \eqref{eq:SNP} is given by $Q/\bar{N}$. When the moment variable $N$ becomes very large, $\zeta_f^{\rm TTMD}$ goes down to the nonperturbative scale $Q_0^2$, which violates the power counting, $Q\gg Q(1- \hat \tau) \gg q_T$, in the factorization theorem. Therefore, we need to freeze the Collins-Soper scale as $\zeta_{N,f}<Q_0^2$

\begin{align}\label{eq:zeta_star}
    \zeta_{*}  \equiv \zeta_{*}(\zeta_{N,f},Q_0) =  \left(\frac{Q}{\bar N}\right)^2\left(1 + \frac{Q_0^2 \bar N^2}{Q^2}\right),
\end{align}
which reproduces the resummation formula in the moment variable $N$ as long as $Q/\bar N \gg Q_0$. Besides,  it also allows us to perform a straightforward numerical calculation in the complex-$N$ plane. We leave the investigation of different function forms and corresponding theoretical uncertainties for future studies. 

\begin{figure}[t]
	\centering
	\includegraphics[width=0.35\textwidth,clip]{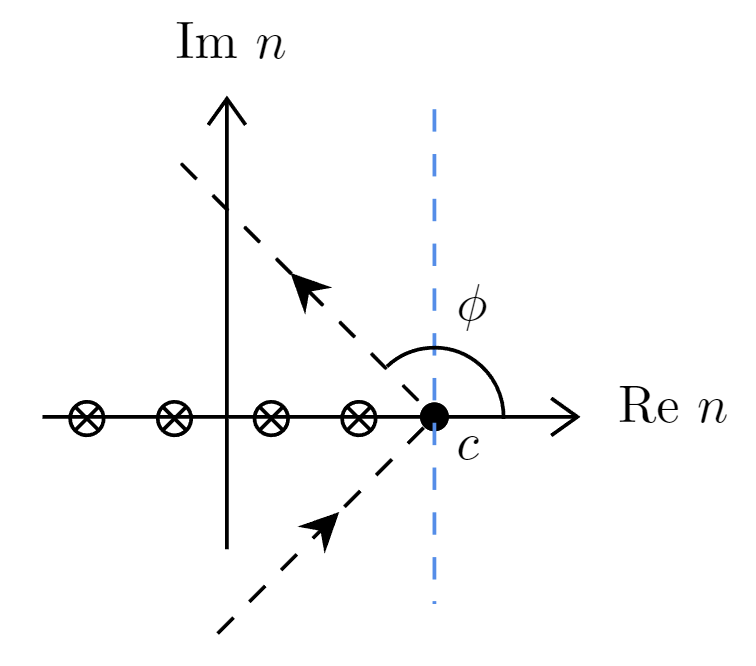} \\
	\caption{Mellin inversion contour. We choose $\phi = \pi/2$ and $c = 1.6$ to adapt various $Q$ in our numerical analysis as shown in the blue dashed line.}\label{fig:Mellin}
\end{figure}

To implement the inverse Mellin transformation, one often rewrites the standard textbook Mellin inversion \cite{Graudenz:1995sk} as an integration over a real variable with adjustable contour: 
\begin{equation}
    \varphi ( x) =\frac{1}{\pi }\int\nolimits _{0}^{\infty }\mathrm{d} z\ \mathrm{Im}\left[\mathrm{exp}( i\phi ) x^{-c-z \exp (i\phi )} \varphi _{n=c+z \exp( i\phi )}\right] .
\end{equation}
Then contour is characterized by a real number $c$, which should be the right to the rightmost singularity of $\varphi _{n}$, and an arbitrary angle $\phi $ as shown in Figure \ref{fig:Mellin}. The parametrization forms we chose are certain simple combinations of power functions, so they only have poles on the real axis. In this regard, one can always find a suitable $c$ and may try various $\phi $ to reach higher numerical efficiency. There are different Mellin inversion prescriptions available in the literature \cite{Catani:1996yz, Graudenz:1995sk}.
However, when we modify the Collins-Soper scale term from $\zeta_f$ to $\zeta_*$, we find that this modified term would contribute two new poles
\begin{equation}
    n=-c\pm i\frac{\bar{Q}}{Q_{0}} ,
\end{equation}
with $\bar{Q} \equiv e^{\gamma _{E}} Q.$ These two singularities would cause numerical issues even when the contour might not exactly pass the poles for different $\bar{Q} /Q_{0}$. For simplicity, we chose $\phi =\pi /2$ for our contour, which is safe no matter what the $Q$ value is. Also, we choose $c=1.6$ to avoid all the other poles. Numerically, we vary the value of $\phi$ and $c$ and find the results are stable and do not depend, within errors, on the particular choice of input parameters.

\begin{figure}[t]
	\centering
	\includegraphics[width=1.0\textwidth,clip]{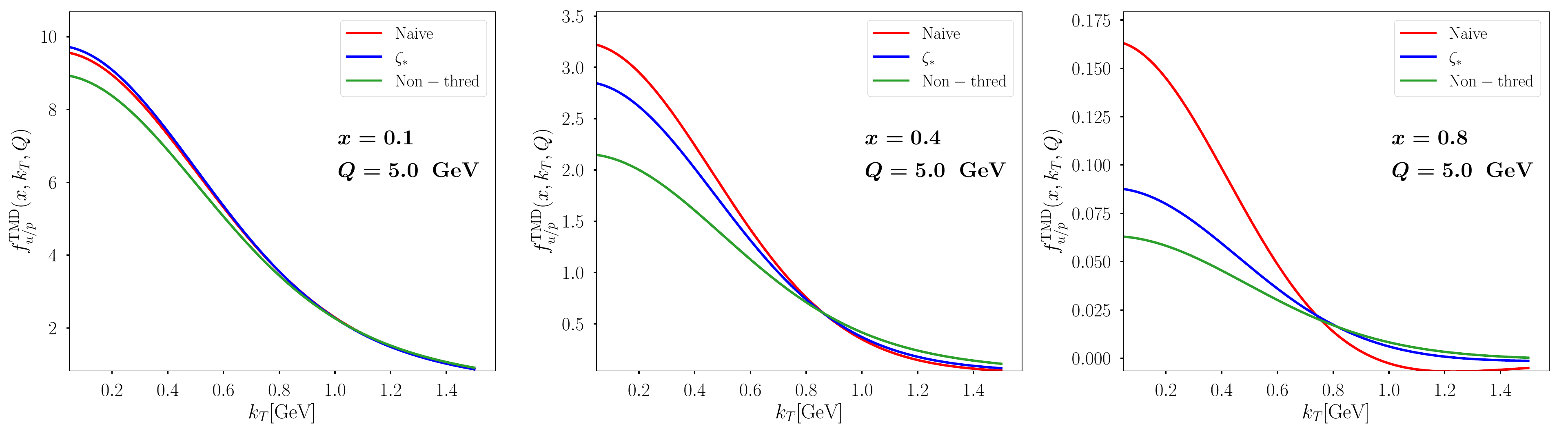} 
	\caption{Transverse momentum distribution of TMD PDFs for the up ($u$) quark in the proton, obtained for three different Bjorken scales $x=0.1, 0.4, 0.8$ at $Q=5$ GeV. The green curves correspond to the results obtained without threshold resummation. The red and blue curves represent the results obtained with threshold resummation using the $\zeta_*$ scheme and without it, respectively. The collinear PDF sets used in this analysis are parameterized according to the {\tt CT18NNLO} prescription, with the parameterized equations specified at the initial scale $\mu_F = 1.3$ GeV.}\label{fig:TMDPDF_TTMDPDF}
\end{figure}

\subsection{Threshold-TMD PDFs and FFs}

In this subsection, we present the numerical results for threshold-improved TMD PDFs and FFs. To investigate the impact of threshold resummation, we compare the TMD PDFs with and without threshold resummation in Figure~\ref{fig:TMDPDF_TTMDPDF}, which shows the transverse momentum distributions of TMD PDFs for the up ($u$) quark in the proton. We vary the value of $x$ from $0.1$ to $0.8$ at $5$ GeV. The green curves represent the results obtained without threshold resummation, corresponding to the case where $\zeta_{N,f}=Q^2$ in \eqref{eq:thrshold_TMDPDF}. The red and blue curves correspond to the results obtained with threshold resummation using the $\zeta_*$ scheme \eqref{eq:zeta_star} and 
naive replacement $\zeta_{N,f}=Q^2/\bar N^2$, respectively. In the low $x$ region, all three curves exhibit consistent behavior since the threshold effect is mild. However, as we increase the value of $x$, the threshold effects become evident. Especially, it is noteworthy that in the naive scheme the theoretical predictions for the distribution rapidly decrease to zero and then turn negative. Consequently, we conclude that the $\zeta_*$ prescription is necessary to ensure reliable theoretical predictions.

\begin{figure}[t]
	\centering
	\includegraphics[width=0.9\textwidth,clip]{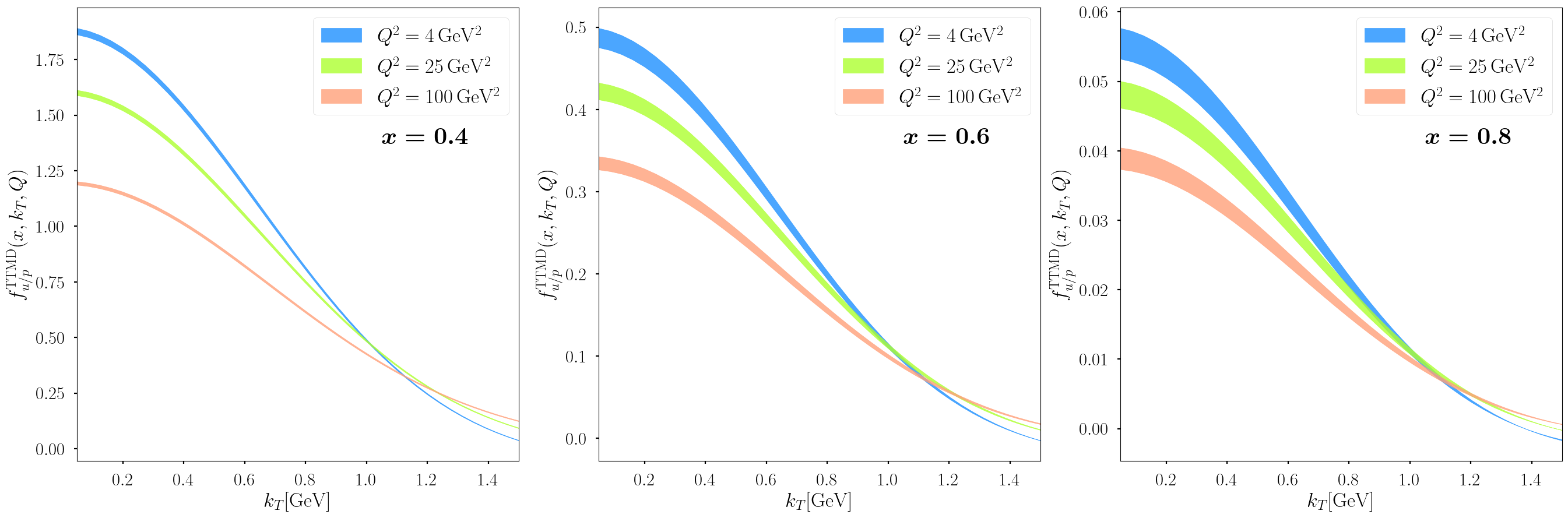} \\
	\vspace{0.5cm}
	\includegraphics[width=0.9\textwidth,clip]{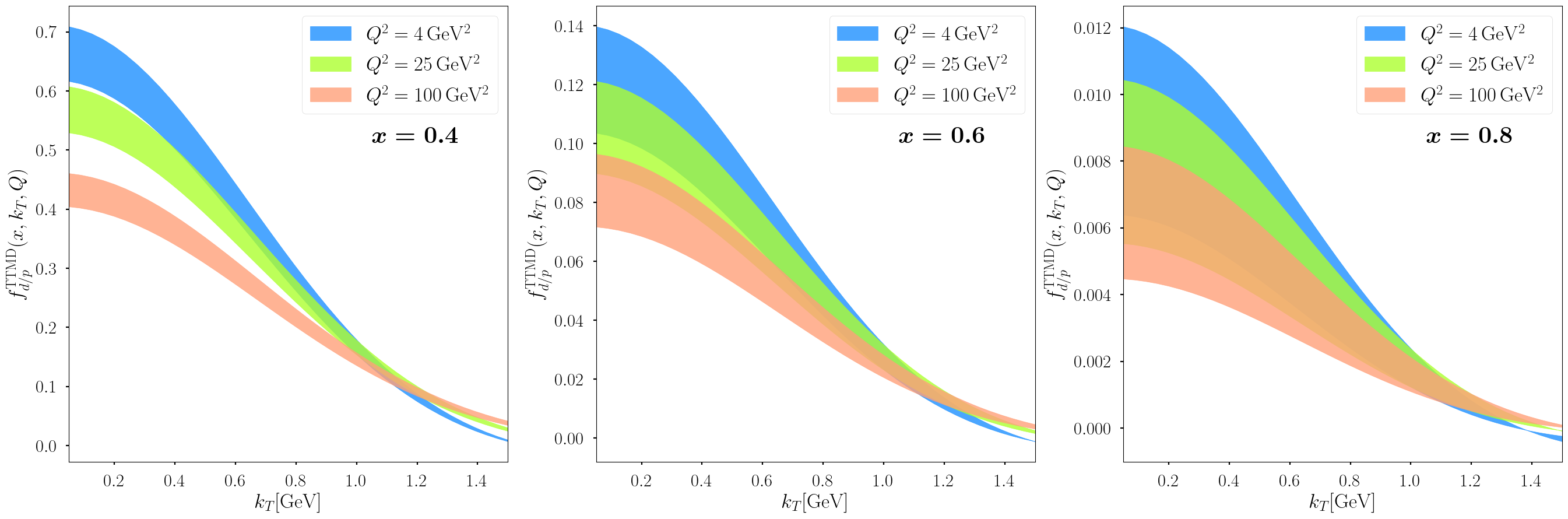}
	\caption{Transverse momentum distribution of threshold-TMD PDFs for $u$ (upper panel) and $d$ (lower panel) quark in the proton for three different hard scales and three different Bjorken scales. Here, we choose parameterized {\tt CT18NNLO} collinear PDF sets where the parameterized equations are given at initial scale $\mu_F = 1.3$ GeV.}\label{fig:TTMDPDF}
\end{figure}

In Figure~\ref{fig:TTMDPDF}, we present the threshold improved TMD PDF for $u$ and $d$ quarks. We make the use of {\tt CT18NNLO} \cite{Hou:2019efy} parameterized colinear PDF set at a given initial scale $\mu_F = 1.3$ GeV. The upper panel is for the $u$ quark and the bottom panel is for the $d$ quark.  
We present our results for three different choices of hard scales; $Q = 2.0, 5.0, 10.0$ GeV with three different choices of Bjorken scales $x = 0.4, 0.6, 0.8$, from left to the right respectively. In the threshold region when the hadronic threshold variable is in the order of $1$, the Bjorken scales are also in the order of $1$. We present our results for a large Bjorken scale to highlight the threshold effect in this region. The uncertainty bands correspond to the 1-$\sigma$ variation from {\tt CT18NNLO} PDFs using the Hessian method. The details of these uncertainties are given in appendix \ref{app:PDF_FF}.

\begin{figure}[t]
	\centering
	\includegraphics[width=0.9\textwidth,clip]{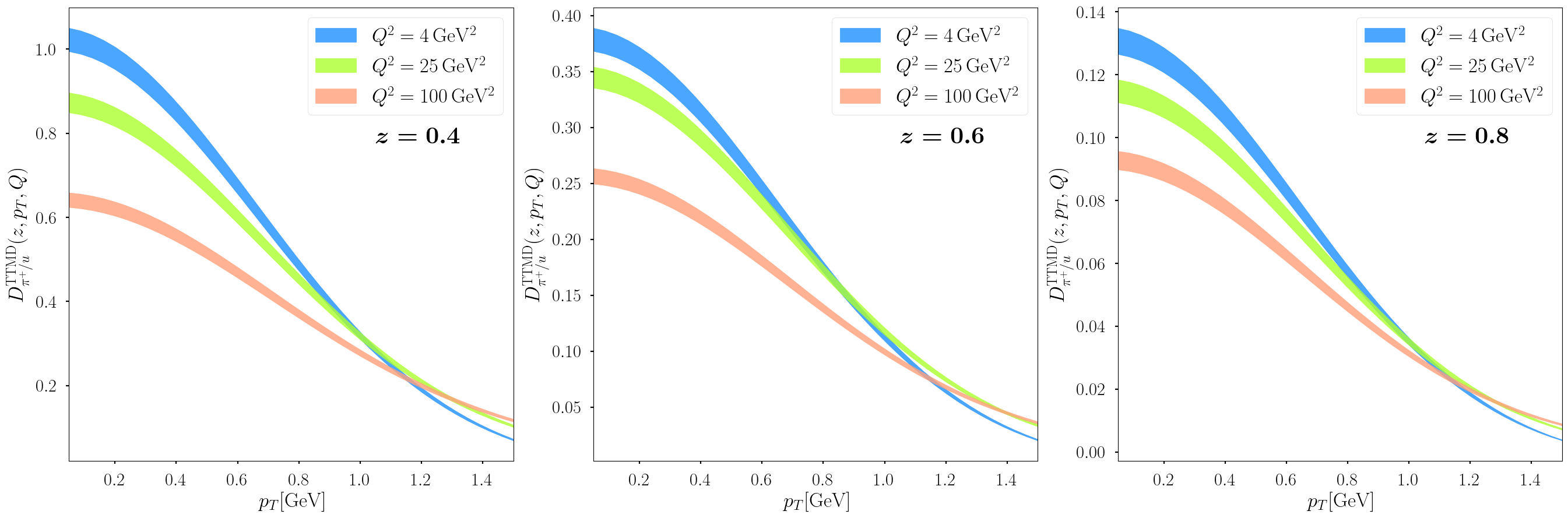} \\
	\vspace{1cm}
	\includegraphics[width=0.9\textwidth,clip]{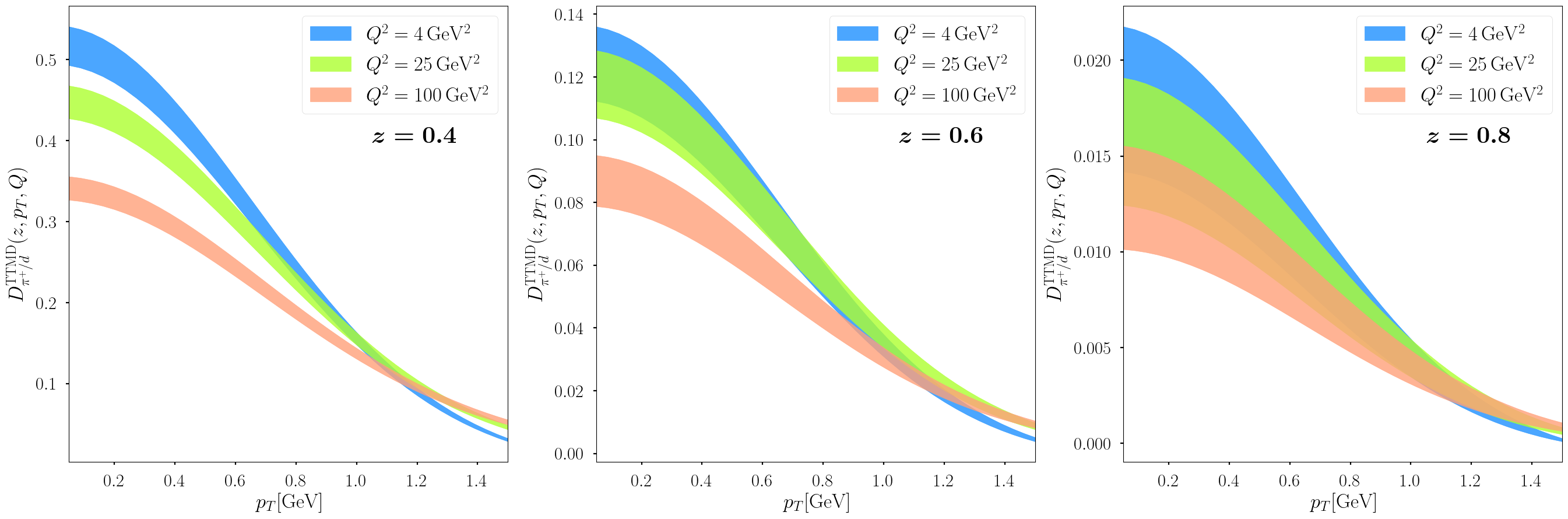}
	\caption{Transverse momentum distribution of threshold-TMD FFs for the $\pi^+$ from  $u$ (upper panel) and $d$ (lower panel) quark. Here, we choose parameterized {\tt JAM20} collinear FF sets at a given scale $\mu_F = 1.3$ GeV.}\label{fig:TTMDFF}
\end{figure}

In Figure~\ref{fig:TTMDFF}, we present threshold improved TMD fragmentation function of Pion ($\pi^{+}$ hadron). In this case, we use the parameterized {\tt JAM20} fragmentation function \cite{Moffat:2021dji} at the scale $\mu_F = 1.3$ GeV. Similar to Figure~\ref{fig:TTMDPDF}, the upper panel is for the up-type and bottom panel is for the down-type quark fragmentation functions and $z$ is the fragmentation variable. Here, we present our results for three different values of hard scale ($Q$) with three different values of fragmentation ($z$) variables. The uncertainty bands correspond to 1-$\sigma$ variation of {\tt JAM} FFs using the replica method.

At this point, we would like to emphasize that the direct comparison between the usual TMD functions and threshold-improved TMD functions is not trivial. In our theoretical formalism, we evolve the TMD evolution factor from the initial scale $\mu_F$ to the desired final scale. On the other hand, in the standard TMD evolution, one starts from scale $\mu_b^*$. Besides, both perturbative and nonperturbative parts in the Collins-Soper evolution are also different in these two frameworks. Because of this different evolution scheme, it is not straightforward to compare them and we leave such a study for future investigation.

\subsection{Predictions for different experiments}

\begin{figure}[t]
	\centering
	\includegraphics[width=0.49\textwidth,clip]{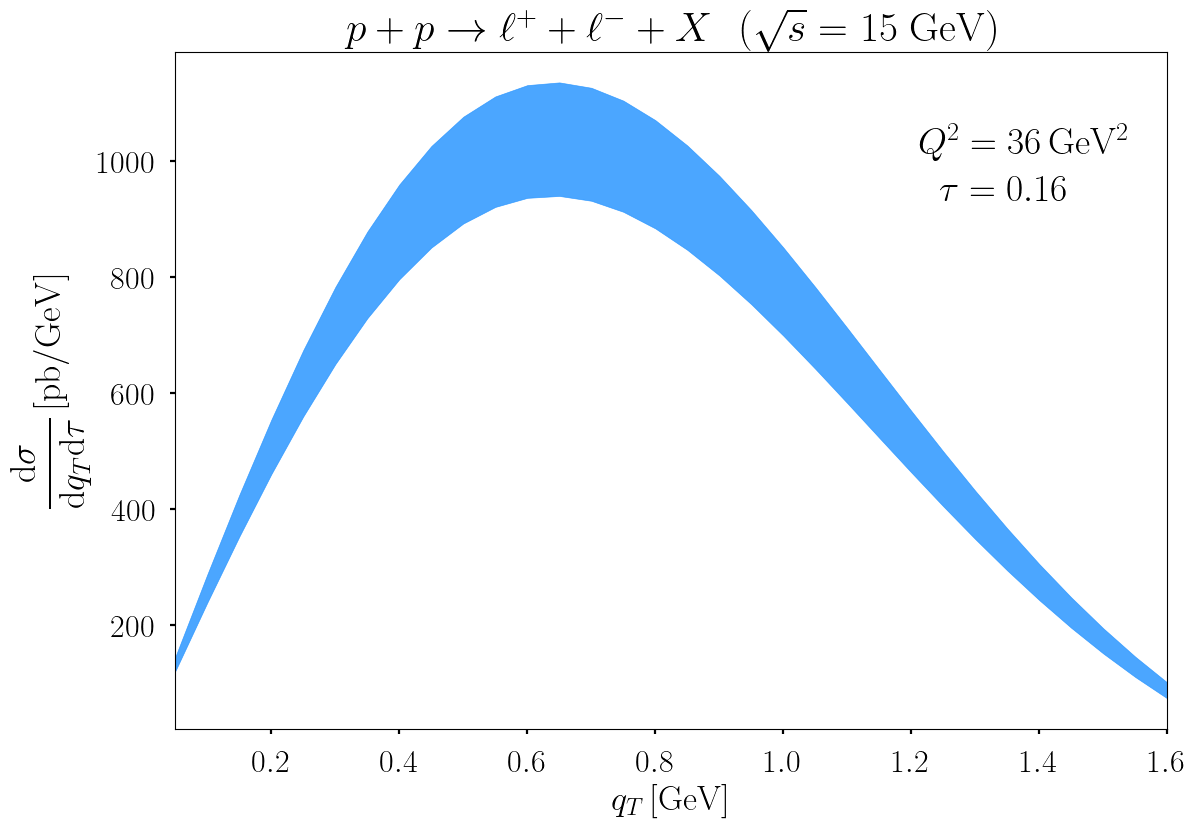}
	\includegraphics[width=0.48\textwidth,clip]{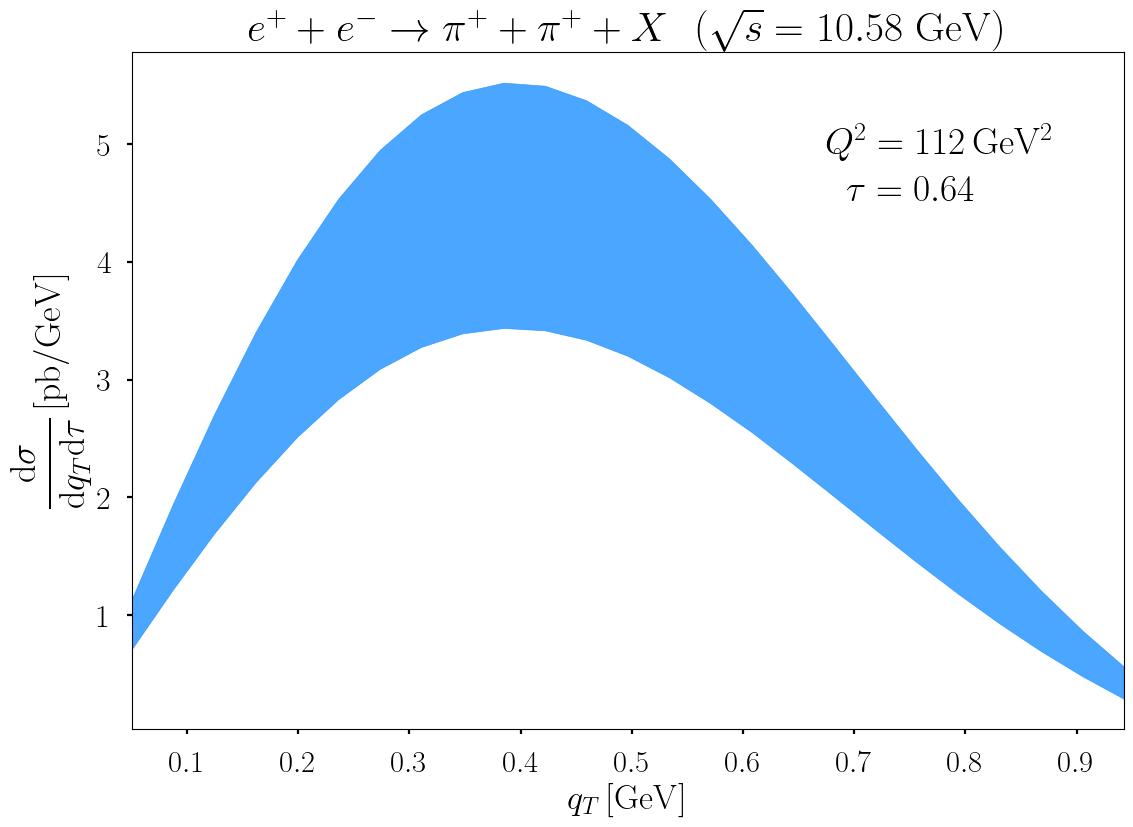} 
\caption{Cross section distribution for Drell-Yan and $e^+e^-$ processes. The left panel is for Drell-Yan dilepton production in $pp$ collisions, $p+p\to \gamma^*\to \ell^++\ell^-+X$. The right panel is for back-to-back two hadron production in $e^+e^-$ collisions, $e^+ + e^-\to \pi^+ + \pi^+ + X$, where we choose both hadrons to be $\pi^+$ as an example. The $e^+e^-$ annihilation result is presented for BELLE kinematics. All the parameters for the numerical computations are mentioned in the main section of the numerical results.}
\label{fig:dy-ee}
\end{figure}

Finally, we use previously determined threshold-improved TMD functions (PDFs and FFs) to predict some experimental results. For this, we investigate three different processes namely, DY, SIDIS, and $e^+e^-$ annihilation process. For the DY process, we consider dilepton transverse momentum distribution from proton-proton collision. We choose the hadronic center of mass energy to be $\sqrt{s} = 15.0$ GeV and $Q = 6.0$ GeV, which is relevant to the Drell-Yan production at typical Fermilab experiments. The results are presented in the left panel of Figure \ref{fig:dy-ee}. For the $e^+e^-$ annihilation process, we consider Pion ($\pi^{+}$) in the final hadronic state. We choose the virtuality of the photon to be $Q = 10.58$ GeV and the threshold variable $\tau = 0.64$ for the BELLE \cite{Belle:2008fdv} kinematics. The results are presented in the right panel of Figure \ref{fig:dy-ee}. The uncertainties in all the cross sections are coming from the threshold-TMD functions which are computed from the 1-$\sigma$ variation of PDF and FF using Hessian and Replica methods respectively. 
As shown in the previous subsection, the uncertainty from FFs is larger than that from PDFs at a similar value of the light-cone momentum fraction making a wider uncertainty band in the $e^+e^-$ process than in the Drell-Yan process.

\begin{figure}[t]	
        \centering
	\includegraphics[width=0.49\textwidth,clip]{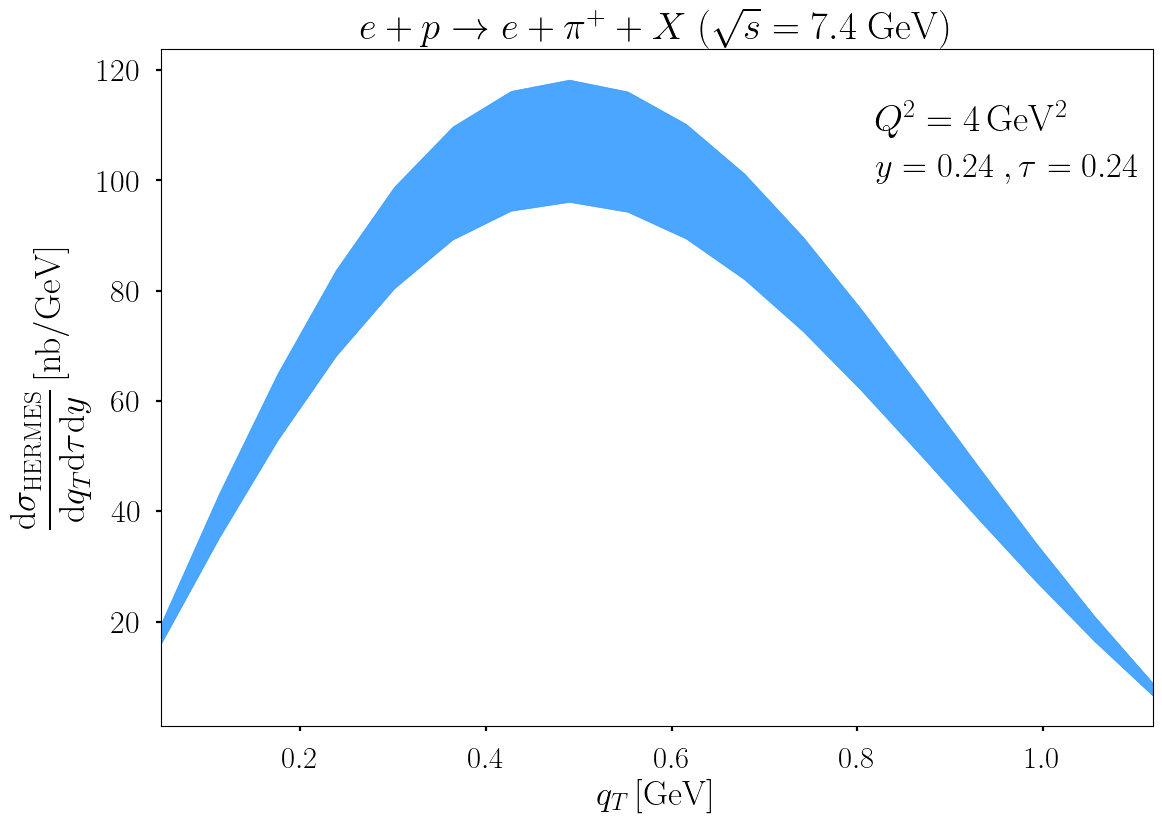}
	\includegraphics[width=0.485\textwidth,clip]{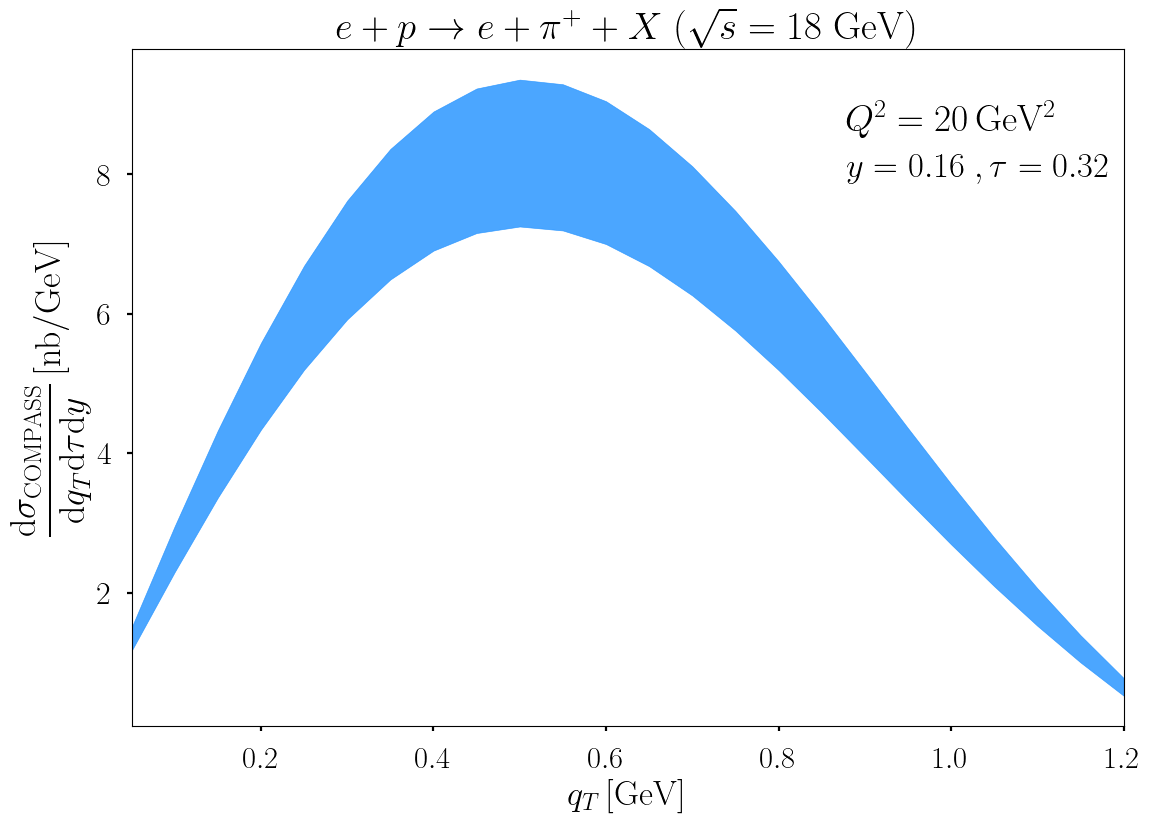} \\
	\vspace{0.5cm}
	\includegraphics[width=0.49\textwidth,clip]{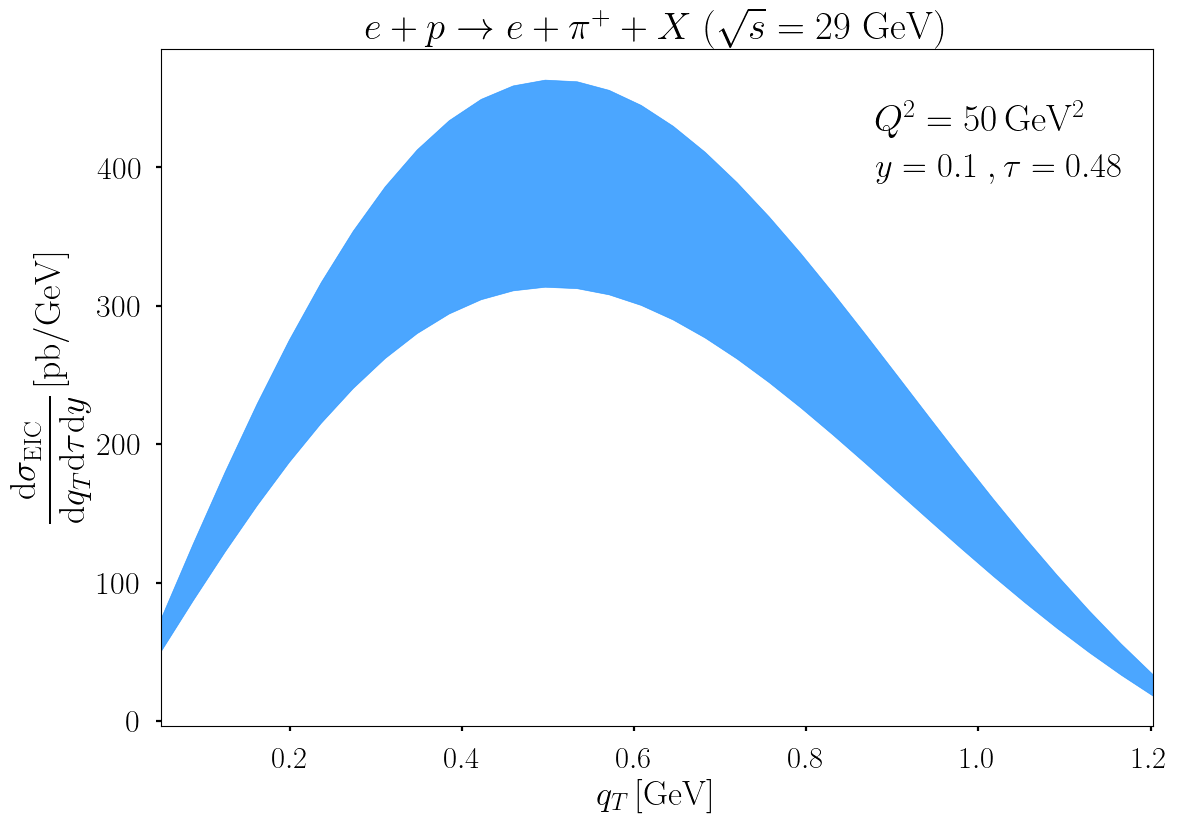} \includegraphics[width=0.48\textwidth,clip]{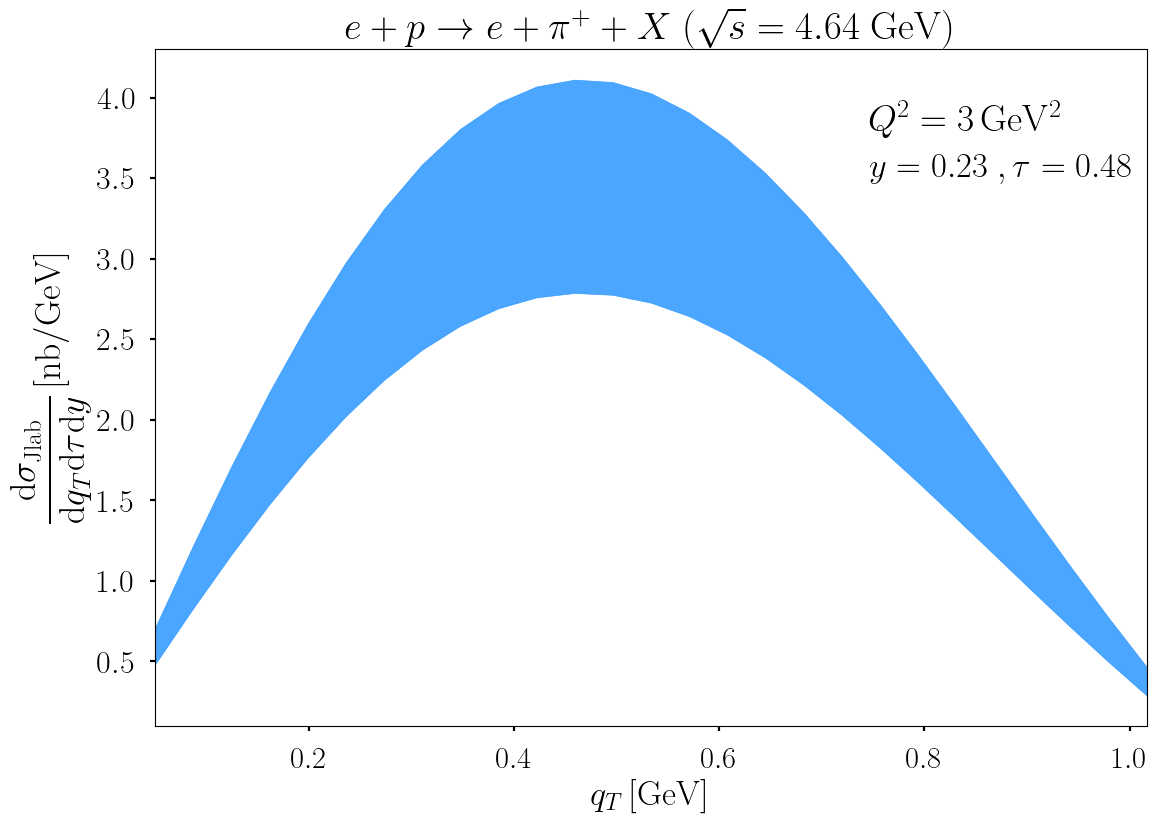}
	\caption{Cross section distribution for SIDIS process, $e+p\to e+\pi^+ + X$, where we choose the final state hadron to be a $\pi^+$ meson as an example. The upper panel is for HERMES (left) and COMPASS (right) kinematics, and the lower panel is for future EIC (left, $5\times 41$ beam energies for $ep$ collisions with $\sqrt{s}=29$ GeV) and JLab 12 GeV program (right). }\label{fig:xsection}
\end{figure}

Finally, in Figure \ref{fig:xsection}, we present the results for the SIDIS process. In this case, we make use of four experiments kinematics namely HERMES \cite{HERMES:2004mhh}, COMPASS \cite{COMPASS:2012ozz}, the future Electron-Ion Collider (EIC) \cite{AbdulKhalek:2021gbh} at Brookhaven National Laboratory, and the $12$ GeV program~\cite{Dudek:2012vr} currently underway at Jefferson Lab (JLab). The upper left panel is for HERMES kinematics of $ep$ collision with hard scale $Q^2 = 4$ GeV$^2$ and threshold parameter $\tau = 0.24$ and the upper right panel is for COMPASS kinematics with hard scale $Q^2 = 20$ GeV$^2$ and threshold parameter $\tau = 0.32$. In the lower panel, we present the results for EIC (left, $Q^2=50$ GeV$^2$ and $\tau=0.4$) and JLab 12 GeV program (right, $Q^2=3$ GeV$^2$ and $\tau=0.48$) kinematics.  To conclude this section, we notice that although the future EIC will make the most precise SIDIS measurements at small $x$ (down to $x\sim 10^{-4}$), the EIC will also increase the precision of the data in the large-$x$ region up to $x \sim 0.5$. The JLab 12 GeV program will make precision measurements up to $x \sim 0.6$ and smaller values of $Q^2$. We expect both EIC and JLab 12 to make important constraints on these threshold TMD PDFs and TMD FFs in the near future.

\section{Conclusion}\label{sec:conclusion}

In this paper, we have introduced the threshold TMD functions namely, PDFs and FFs. To probe this threshold effect, we use three different processes: Drell-Yan, SIDIS and $e^+e^-$ processes. Considering the extra degrees of freedom in this kinematical region known as collinear-soft degrees of freedom, we re-factorized the formula for these processes and defined threshold-improved TMD distribution functions. We observe that because of the phase space reduction in the threshold region, the Collins-Soper scale is not the same for the usual TMD and threshold-improved TMD distribution functions. We find that $\zeta_{N,f} = Q^2/\bar{N}^2$ for threshold TMD whereas $\zeta_f=Q^2$ for the usual TMD. During our numerical analysis, we observe that apart from the RG consistency, the Collins-Soper scale has to be chosen in such a way that it is always greater than the non-perturbative scale $Q_0$. Therefore, one needs to modify the Collins-Soper scale, and this modified scale introduces a new kind of pole in the integration contour. We provide a Mellin inversion prescription to avoid all kinds of poles in the integration contour. Finally, we provide the numerical predictions for transverse momentum distributions of Drell-Yan, SIDIS and $e^+e^-$ processes for different experimental kinematics using these threshold-improved TMD functions.

Our theoretical formalism would open a new window into TMD physics. Future experimental analysis and global fitting analysis will certainly help in understanding the non-perturbative mechanism of the TMD functions in the threshold limit and unveiling the three-dimensional picture of a hadron in the large $x$ limit. Our formalism will be important to extract these TMD functions in the threshold region and will be reliable theoretical input to understand the experimental data. In this work, we only present the factorization and resummation formula for the unpolarized cross section, but it can be generalized to the process involving polarized hadron in both initial and final states. The corresponding theoretical predictions on the spin asymmetry in the threshold limit will be explored in future work.

\acknowledgments
We thank Hui-Chun Chin, Jun Gao, Dianyu Liu, Zekun Lou and Yiyu Zhou for their useful discussions. Z.K. is supported by the National Science Foundation under grant No.~PHY-1945471. K.S. and D.Y.S. are supported by the National Science Foundations of China under Grant No.~12275052 and No.~12147101 and the Shanghai Natural Science Foundation under Grant No.~21ZR1406100. Y.Z. is supported by the Tsung-Dao Lee Chinese Undergraduate Research Endowment under Grant No.~202210246155.

\appendix
\section{Collinear-soft function}\label{app:csfun}
In this appendix, we furnish a one-loop perturbative analysis of the collinear-soft functions. Additionally, we investigate the threshold asymptotics of the perturbative matching coefficients for TMD PDFs, elucidating their equivalence to the collinear-soft functions. We then extend our analysis to offer comprehensive expressions for the perturbative expansion of the collinear-soft function, valid up to three-loop order. Importantly, we corroborate that these expressions are consistent with the Collins-Soper equations, paralleling the behavior of standard TMD PDFs. Furthermore, we confirm the RG invariance up to the third loop level.

In the following calculation, the light-like basis vectors are defined as 
\begin{align}
    n^\mu = (1,0,0,1),~~~~\bar n^\mu = (1,0,0,-1).
\end{align}
In order to regularize the rapidity divergences we add the factor 
\begin{align}
    \int \mathrm{d}^d k \to \int \mathrm{d}^d k \left(\frac{\nu}{2|k_z|}\right)^\eta
\end{align}
in the phase space integrals \cite{Chiu:2011qc,Becher:2010tm} in the collinear-soft function. This regulator preserves the $n^\mu \leftrightarrow \bar n^\mu$ symmetry of the process. Explicitly, the momentum scaling of the collinear-soft modes $k_{cs}^\mu$ is 
\begin{align}
   k^\mu_{cs}\equiv (\bar n\cdot k_{cs}, \, n \cdot k_{cs} , \, k_{cs,\perp})\sim\left(Q(1-\hat \tau), \frac{q_T^2}{Q(1-\hat \tau)}, q_T\right),
\end{align}
so the rapidity regulator should be further expanded as $(\nu/\bar n \cdot k)^\eta$ at the leading power due to $q_T \ll Q(1 - \hat \tau)$. Similarly, the rapidity regulator is expressed as $(\nu/n \cdot k)^\eta$ in the anti-collinear-soft sector. Then the one-loop unsubtracted bare collinear-soft function in $N$-space is defined as 
\begin{align}\label{eq:oneloop-unsub-csoft}
    \tilde S_c^{0,{\rm unsub}}(b_T,\epsilon,\eta,Q/\bar{N}) = \,&1 + C_F g_s^2 \left(\frac{\mu^2 e^{\gamma_E}}{4\pi}\right)^\epsilon  \int \frac{\mathrm{d}^d k}{(2\pi)^{d-1}}\delta(k^2)\theta(k^0) \left(\frac{\nu}{\bar n\cdot k}\right)^\eta \frac{2n\cdot \bar n}{n\cdot k\, k\cdot \bar n} \notag \\
    & \times e^{i \bm{k}_T \cdot \bm{b}_T} e^{-N \bar n \cdot k/Q} \notag \\
    = \,& 1 + \frac{\alpha_s}{4\pi}C_F \left[\left(\frac{2}{\eta} + \ln \frac{\nu^2\bar N^2}{Q^2}\right)\left(\frac{2}{\epsilon} + 2 L_b\right) \right],
\end{align}
where we apply the approximation $ (1-z)^N \approx  e^{-Nz} $ after expanding out the power suppressed terms in the large $N$ limit. We note that the rapidity divergence is the same as the one in TMD PDFs, which should be expected based on the factorization formula. After factorizing the soft function, one can write
\begin{align}
    \label{eq:c-soft}
    \tilde{S}_c(b_T,\mu,\zeta_N) &= \lim _{\substack{\epsilon \rightarrow 0 \\ \eta \rightarrow 0}} Z_{\mathrm{uv}}(\mu, \zeta_N, \epsilon) \tilde S_c^{0,{\rm unsub}}(b_T,\epsilon,\eta,Q/\bar{N}) \sqrt{S^0(b_T,\epsilon,\eta)} \notag \\
    &=\tilde{S}_c^{\text{unsub}}(b_T,\mu,\zeta_N/\nu^2) \sqrt{S(b_T,\mu,\nu)},
\end{align}
where the one-loop bare soft function reads~\cite{Chiu:2012ir,Kang:2017glf}
\begin{equation}
    S^0(b_T,\epsilon,\eta) = 1 + \frac{\alpha_s}{4\pi} C_F \left[ -\left(\frac{2}{\eta} + \ln \frac{\nu^2 }{\mu^2}\right)\left(\frac{4}{\epsilon}+ 4 L_b \right) + \frac{4}{\epsilon^2} -2 L_b^2 - \frac{\pi^2}{3} \right].
\end{equation}
 In addition, we replaced $Q^2/\bar N^2$ with the Collins-Soper scale $\zeta_N$. Therefore, at one-loop order \eqref{eq:c-soft} has the form as 
\begin{align}\label{eq:oneloop-csoft}
\tilde{S}_c(b_T,\mu,\zeta_N)= 1 + \frac{\alpha_s}{4\pi} C_F \left(-L_b^2 + 2L_b \ln \frac{\mu^2}{\zeta_N} -\frac{\pi^2}{6} \right), 
\end{align}
where rapidity divergences related to the poles in $\eta$ cancel out during the combination and the remainder divergences in $\epsilon$ are absorbed into the renormalization factor $Z_{\mathrm{uv}}$ by defining a finite collinear-soft function based on the standard RG methods.  

Finally, we want to point out that the perturbative results of the collinear-soft function can also be obtained by taking the $N\to\infty$ limit in the TMD PDFs or TMD FFs \cite{Echevarria:2016scs}. In the following discussion, we use TMD PDFs as our example. After performing operator product expansion, the $x$-space unsubtracted TMD PDF can be expressed as
\begin{equation}
    f^{\rm unsub}_{q/h}(x,b_T,\mu,\zeta/\nu^2) = \sum_{i} \int_{x}^{1} \frac{dz}{z} C_{q\leftarrow i}(z,b_T,\mu,\zeta/\nu^2) f_{i/h}(x/z,\mu) + \mathcal{O}(b_T^2\Lambda_{\text{QCD}}^2),
\end{equation}
where $C_{q\leftarrow i}$ is the perturbative matching coefficient, and in the moment space at one loop it can be written as 
\begin{align}
    C_{q\leftarrow q} (z,b_T,\,& \mu,\zeta/\nu^2) =  \delta(1-z)  \\
    & + \frac{\alpha_s}{4\pi} \left[C_FL_b\bigg(3+2\ln \frac{\nu^2}{\zeta} \bigg)\delta(1-z) + \frac{P^{(0)}_{q\leftarrow q}(z)}{2}L_b + 2C_F(1-z) \right] + \mathcal{O}(\alpha_s^2), \notag
\end{align}
where 
\begin{align}
    P^{(0)}_{q \leftarrow q}(z) = 4C_{F}\Bigg[\frac{1+z^{2}}{( 1-z)_{+}} +\frac{3}{2} \delta ( 1-z)\Bigg]
\end{align}
is the one-loop Altarelli-Parisi splitting kernel. In the Mellin $N$-space the matching coefficient can be written as
\begin{align}
\tilde{C}_{q \leftarrow q}\left(N, b_T, \mu, \zeta / \nu^{2}\right) & =1 \\
& \hspace{-1.5cm}+ \frac{\alpha_{s}}{4 \pi}\left[C_{F} L_{b}\left(3+2 \ln \frac{\nu^{2}}{\zeta}\right)-C_{F} L_{b}(-4 \ln \bar{N}+3)+2 C_{F} \frac{1}{N(N+1)}\right]+\mathcal{O}\left(\alpha_{s}^{2}\right). \notag
\end{align}
Then we take the large $N$ limit and can have 
\begin{align}
    \tilde{C}_{q \leftarrow q}\left(N, b_T, \mu, \zeta / \nu^{2}\right) \xrightarrow{N \rightarrow \infty} \tilde{S}_{c}^{\text {unsub }}(b_T,\mu,\zeta_N/\nu^2) =1+\frac{\alpha_{s}}{4 \pi} C_{F}\left[2 L_{b} \ln \frac{\nu^{2}}{\zeta_N}\right],
\end{align}
where the contribution from the flavor off-diagonal splitting kernels is power suppressed in the threshold limit, and so can be neglected in the leading-power factorization formula.  

To rigorously validate the factorization theorem in the joint threshold and TMD limit, examining higher-order corrections in the perturbative expansion of collinear-soft functions should be illuminating. Specifically, we employ the threshold asymptotics of the NNNLO expressions for the perturbative matching coefficients $C_{q\leftarrow q}$ \cite{Luo:2019szz,Luo:2020epw,Ebert:2020yqt} to ascertain the three-loop collinear-soft function. We represent the expansion coefficients of the unsubtracted collinear-soft functions as
\begin{align}\label{eq:cs-threeloop}
     \tilde{S}_{c}^{\text {unsub }} (b_T,\mu,\zeta_N/\nu^2)  = 1 + \sum_{n=1}^\infty \left(\frac{\alpha_s}{4\pi}\right)^n \tilde{S}_{c}^{(n)},
\end{align} 
where the coefficients up to three loop are given by
\begin{align}
    \tilde{S}_{c}^{(1)} = & \, 2 C_{F} L_{b} L_\zeta, \notag \\
    \tilde{S}_{c}^{(2)} = &
\, 2 C_F^2 L_b^2 L_\zeta^2 + C_F C_A \left[ \frac{11}{3} L_b^2  + \left( \frac{134}{9} - 4 \zeta_2 \right) L_b + \left( \frac{404}{27} - 14 \zeta_3 \right) \right] L_\zeta \notag\\
  & + C_F T_F n_f \left( - \frac{4}{3} L_b^2 - \frac{40}{9} L_b - \frac{112}{27}      \right) L_\zeta, \notag \\
    \tilde{S}_{c}^{(3)} = & \, \frac{4}{3} C_F^3 L_b^3 L_\zeta^3  + C_F^2 C_A  \left[ \frac{22}{3} L_b^3  + \left( \frac{268}{9} - 8 \zeta_2 \right) L_b^2  + \left( \frac{808}{27} - 28 \zeta_3 \right) L_b  \right] L_\zeta^2 \notag\\
    & + C_F C_A^2  \left[ \frac{242}{27} L_b^3  + \left( \frac{1780}{27} - \frac{44}{3} \zeta_2 \right) L_b^2  + \left( \frac{15503}{81} - \frac{536}{9} \zeta_2 - 88 \zeta_3 + 44 \zeta_4 \right) L_b  \right. \notag\\
    & \left. + \left( \frac{297029}{1458} - \frac{3196}{81} \zeta_2 - \frac{6164}{27} \zeta_3 + \frac{88}{3} \zeta_2 \zeta_3 - \frac{77}{3} \zeta_4 + 96 \zeta_5 \right) \right] L_\zeta \notag\\
    & + C_F C_A T_F n_f  \left[ -\frac{176}{27} L_b^3 + \left( -\frac{1156}{27} + \frac{16}{3} \zeta_2 \right) L_b^2  + \left( -\frac{8204}{81} + \frac{160}{9} \zeta_2 \right) L_b  \right. \notag\\
    & \left. + \left( -\frac{62626}{729} + \frac{824}{81} \zeta_2 + \frac{904}{27} \zeta_3 - \frac{20}{3} \zeta_4 \right)  \right] L_\zeta \notag\\
    &+ C_F T_F^2 n_f^2  \left( \frac{32}{27} L_b^3 + \frac{160}{27} L_b^2+ \frac{800}{81} L_b     + \frac{3712}{729}  + \frac{64}{9} \zeta_3  \right)L_\zeta \notag\\
    & + C_F^2 T_F n_f  \left[ -\frac{8}{3} L_b^3 L_\zeta^2 + \left( -4 - \frac{80}{9} L_\zeta \right) L_b^2 L_\zeta + \left( -\frac{224}{27}  -\frac{110}{3} + 32 \zeta_3  \right) L_b L_\zeta^2 \right. \notag\\
    & \left. + \left( -\frac{1711}{27} + \frac{304}{9} \zeta_3 + 16 \zeta_4 \right) L_\zeta \right] ,
\end{align}
where $L_\zeta\equiv\ln(\nu^2/\zeta_N)$. We can immediately confirm that the above expressions satisfy the standard Collins-Soper equations. Upon factorizing the soft function, as previously outlined in Eq.\,\eqref{eq:c-soft}, we procure the properly subtracted collinear-soft function  $\tilde{S}_{c} (b_T,\mu,\zeta_N)$. Furthermore, the non-cusp anomalous dimensions are as follows:
\begin{align}
    \gamma_0^{\tilde S_c} &= 0, \notag \\
    \gamma_1^{\tilde S_c} &= C_F C_A \left(\frac{404}{27} - \frac{11 \pi^2}{18} - 14 \zeta_3\right) + C_F T_F n_f \left(-\frac{112}{27} + \frac{2 \pi^2}{9}\right) , \notag \\\
    \gamma_2^{\tilde S_c} &=  C_F^2 T_F n_f \left(-\frac{1711}{27} + \frac{2 \pi^2}{3} + \frac{8 \pi^4}{45} + \frac{304 }{9}\zeta_3\right) \notag \\
    &+ C_F C_A^2  \left(\frac{136781}{1458} - \frac{6325 \pi^2}{486} + \frac{44 \pi^4}{45} - \frac{658 }{3}\zeta_3 + \frac{44 \pi^2 }{9}\zeta_3 + 96 \zeta_5\right) \notag \\
    &+ C_F C_A T_F n_f \left(-\frac{11842}{729} + \frac{1414 \pi^2}{243} - \frac{8 \pi^4}{15} + \frac{728 }{27}\zeta_3\right) \notag \\
    &+ C_F T_F^2 n_f^2 \left(-\frac{4160}{729} - \frac{40 \pi^2}{81} + \frac{224 }{27}\zeta_3\right),
\end{align}
which allows us to corroborate the RG consistency condition $\Gamma^h +2 \, \Gamma^{\tilde{S}_c} + 2 \, \Gamma^{\tilde{f}_q}=0$ up to the third-loop level. Below, we also provide, for convenience, the anomalous dimensions for the hard function and the threshold PDF \cite{Becher:2007ty}:
\begin{align}
    \gamma_0^V= &\, -6 C_F, \notag\\
    \gamma_1^V= &\, C_F^2\left(-3+4 \pi^2-48 \zeta_3\right)+C_F C_A\left(-\frac{961}{27}-\frac{11 \pi^2}{3}+52 \zeta_3\right)+C_F n_f T_F \left(\frac{260}{27}+\frac{4 \pi^2}{3}\right), \notag\\
    \gamma_2^V= &\, C_F^3\left(-29-6 \pi^2-\frac{16 \pi^4}{5}-136 \zeta_3+\frac{32 \pi^2}{3} \zeta_3+480 \zeta_5\right) \notag\\
    & +C_F^2 C_A\left(-\frac{151}{2}+\frac{410 \pi^2}{9}+\frac{494 \pi^4}{135}-\frac{1688}{3} \zeta_3-\frac{16 \pi^2}{3} \zeta_3-240 \zeta_5\right) \notag\\
    & +C_F C_A^2\left(-\frac{139345}{1458}-\frac{7163 \pi^2}{243}-\frac{83 \pi^4}{45}+\frac{7052}{9} \zeta_3-\frac{88 \pi^2}{9} \zeta_3-272 \zeta_5\right) \notag\\
    & +C_F^2 T_F n_f\left(\frac{5906}{27}-\frac{52 \pi^2}{9}-\frac{56 \pi^4}{27}+\frac{1024}{9} \zeta_3\right) \notag\\
    & +C_F C_A T_F n_f\left(-\frac{34636}{729}+\frac{5188 \pi^2}{243}+\frac{44 \pi^4}{45}-\frac{3856}{27} \zeta_3\right) \notag\\
    & +C_F T_F^2 n_f^2\left(\frac{19336}{729}-\frac{80 \pi^2}{27}-\frac{64}{27} \zeta_3\right) ,\\
    \gamma_0^{\tilde f_q}= &\, 3 C_F, \notag \\
    \gamma_1^{\tilde f_q}= & \,C_F^2\left(\frac{3}{2}-2 \pi^2+24 \zeta_3\right)+C_F C_A\left(\frac{17}{6}+\frac{22 \pi^2}{9}-12 \zeta_3\right)-C_F T_F n_f\left(\frac{2}{3}+\frac{8 \pi^2}{9}\right), \notag \\
    \gamma_2^{\tilde f_q}= & \, C_F^3\left(\frac{29}{2}+3 \pi^2+\frac{8 \pi^4}{5}+68 \zeta_3-\frac{16 \pi^2}{3} \zeta_3-240 \zeta_5\right) \notag \\
    & +C_F^2 C_A\left(\frac{151}{4}-\frac{205 \pi^2}{9}-\frac{247 \pi^4}{135}+\frac{844}{3} \zeta_3+\frac{8 \pi^2}{3} \zeta_3+120 \zeta_5\right) \notag \\
    & +C_F^2 T_F n_f\left(-46+\frac{20 \pi^2}{9}+\frac{116 \pi^4}{135}-\frac{272}{3} \zeta_3\right) \notag \\
    & +C_F C_A^2\left(-\frac{1657}{36}+\frac{2248 \pi^2}{81}-\frac{\pi^4}{18}-\frac{1552}{9} \zeta_3+40 \zeta_5\right) \notag \\
    & +C_F C_A T_F n_f\left(40-\frac{1336 \pi^2}{81}+\frac{2 \pi^4}{45}+\frac{400}{9} \zeta_3\right) \notag\\
    & +C_F T_F^2 n_f^2\left(-\frac{68}{9}+\frac{160 \pi^2}{81}-\frac{64}{9} \zeta_3\right) .
\end{align}

\section{Fitting parameters for PDFs and FFs}\label{app:PDF_FF}
In this appendix, we will collect parametrization functional forms of PDF and FF at the initial scale $\mu_F$, where we consider {\tt CT18} PDF sets \cite{Hou:2019efy} and {\tt JAM20} FF sets used in section \ref{sec:numerics}. In {\tt CT18} PDF sets \cite{Hou:2019efy}, the valence quarks is parameterized as
\begin{equation}\label{eq:PDFv}
f_{q_v/p}\left(x, \mu_F\right)=a_0 x^{a_1-1}(1-x)^{a_2} P_v(\sqrt{x}),
\end{equation}
with $P_v(y)$ a sum of Bernstein polynomials defined by
\begin{equation}
    P_v(y)=a_3(1-y)^4+4 a_4 y(1-y)^3+6 a_5 y^2(1-y)^2+4\left(1+a_1 / 2\right) y^3(1-y)+y^4 .
\end{equation}
Similarly, for the sea quarks, one has
\begin{equation}\label{eq:PDFsea}
f_{q_{{\rm sea}}/p}\left(x, \mu_F\right)=a_0 x^{a_1-1}(1-x)^{a_2} P_{\text {sea}}\left(1-(1-\sqrt{x})^4\right),
\end{equation}
where
\begin{equation}
    P_{\rm {sea }}(y)=(1-y)^5+5 a_3 y(1-y)^4+10 a_4 y^2(1-y)^3+10 a_5 y^3(1-y)^2+5 a_6 y^4(1-y)+a_7 y^5 .
\end{equation}
The best-fit values of the parameters $a_k$  have been given in \cite{Hou:2019efy}, and we collect them in Table \ref{tab:PDFs}. Besides, we also provide their Hessian uncertainties which have been included in our numerics as the theoretical uncertainty estimation. In order to evaluate the Hessian uncertainties, we apply the following formula
\begin{align}\label{eq:PDFerr}
    \delta^{+} F &=\sqrt{\sum_{i=1}^{N_d}\bigg[\max \left(F_{2 i-1}-F_0, F_{2 i}-F_0, 0\right)\bigg]^2}, \\
    \delta^{-} F &=\sqrt{\sum_{i=1}^{N_d}\bigg[\max \left(F_0-F_{2 i-1}, F_0-F_{2 i}, 0\right)\bigg]^2},
\end{align}
where $F_0$ represents the value of the central PDF of the Hessian set, and $F_{2i-1} (F_{2i})$ represents the value of the error PDF of the Hessian set in the positive (negative) directions of the $i_{\text{th}}$ eigenvectors in the $N_d$ dimensional PDF parameter space. Then we use the functional forms \eqref{eq:PDFv} and \eqref{eq:PDFsea} to fit the uncertainties and get best-fitting parameters. The fitting results are also collected in Table \ref{tab:PDFs} as errors of parameters $a_k$. 

\begin{table}[!t]
\begin{center}
    \newcolumntype{K}{>{\centering\arraybackslash}X}
    \begin{tabularx}{\textwidth}{XKKKKK}
    \toprule\toprule 
    {{ $\text{\tt CT18}$} } & $u_{v}$ & $d_{v}$ & $u_{\text{sea}} =\bar{u}$ & $d_{\text{sea}} =\bar{d}$ & $s=\bar{s}$ \\
    \midrule 
     $a_{0}$ & {\small $3.385_{-0.988}^{+0.066}$} & {\small $0.490_{-0.072}^{+0.670}$} & {\small $0.414_{+0.347}^{-0.012}$} & {\small $0.414_{+0.334}^{-0.037}$} & {\small $0.288_{-0.152}^{+0.111}$} \\
    $a_{1}$ & {\small $0.763_{-0.002}^{+0.005}$} & {\small $0.763_{-0.002}^{-0.005}$} & {\small $-0.022_{+0.084}^{-0.020}$} & {\small $-0.022_{+0.082}^{-0.025}$} & {\small $-0.022_{+0.616}^{-0.022}$} \\
    $a_{2}$ & {\small $3.036_{-0.165}^{-0.024}$} & {\small $3.036_{+0.230}^{+0.301}$} & {\small $7.737_{-0.415}^{+0.152}$} & {\small $7.737_{-0.676}^{+0.870}$} & {\small $10.31_{+0.510}^{+0.170}$} \\
    $a_{3}$ & {\small $1.502_{+1.740}^{+0.611}$} & {\small $2.615_{+6.155}^{-0.107}$} & {\small $0.618_{-0.608}^{+0.191}$} & {\small $0.292_{-0.814}^{+0.517}$} & {\small $0.466_{+1.599}^{+0.029}$} \\
    $a_{4}$ & {\small $-0.147_{-0.465}^{+0.072}$} & {\small $1.828_{+0.406}^{-0.155}$} & {\small $0.195_{+0.525}^{-0.206}$} & {\small $0.647_{+0.888}^{-0.585}$} & {\small $0.466_{+2.482}^{-0.197}$} \\
    $a_{5}$ & {\small $1.671_{+2.319}^{+0.813}$} & {\small $2.721_{+7.609}^{+0.163}$} & {\small $0.871_{-0.906}^{+0.336}$} & {\small $0.474_{-1.238}^{+0.917}$} & $0.255_{-0.992}^{+0.260}$ \\
    $a_{6}$ & ... & ... & {\small $0.267_{+0.305}^{-0.172}$} & {\small $0.741_{+0.443}^{-0.645}$} & {\small $0.255_{+1.772}^{-0.276}$} \\
    $a_{7}$ & ... & ... & {\small $0.733_{-0.595}^{+0.267}$} & {\small $1_{-0.696}^{+0.990}$} & $1_{-1.098}^{+0.035}$ \\
     \bottomrule\bottomrule
    \end{tabularx}
\end{center}\caption{Best-fit parameter values for the proton PDFs of the {\tt CT18} at the initial scale $\mu_F=1.3\,{\rm GeV}$, where the explicit parametrization functional forms are given in \eqref{eq:PDFv} and \eqref{eq:PDFsea} for valence and sea quark, respectively. The errors are the Hessian uncertainties calculated as explained after \eqref{eq:PDFsea}. The upper (lower) part of the numbers is the difference from the central values corresponding to the upper (lower) bound of the PDFs.}\label{tab:PDFs}
\end{table}

For the pion FFs, we apply {\tt JAM20}-SIDIS sets, and the corresponding parametrization functional form reads 
\begin{equation}\label{eq:FFpara}
    D_{\pi^+/q}(z, \mu_F)= \frac{z^{a_0} (1-z)^{a_1}(1+a_2 \sqrt{z} + a_3 z)}{\int\nolimits _{0}^{1}\mathrm{d} z\ z^{a_{1} +1}( 1-z)^{a_{2}}\left( 1+a_{3}\sqrt{z} +a_{4} z\right)},
\end{equation}
with initial scales are $\mu_F=1.3 \, \text{GeV}$. The best fitting parameters are summarized in Table \ref{tab:FFs}, where we  give the error of $a_k$ corresponding uncertainties of FFs evaluated by the replica method. 

\begin{table}
\begin{center}
\newcolumntype{K}{>{\centering\arraybackslash}X}
\begin{tabularx}{\textwidth}{XKK}
\toprule\toprule 
{{\tt JAM20} } & $u=\bar{d}$ & $d=\bar{u} =s=\bar{s}$ \\
\midrule 
$a_{0}$ & $-1.372{_{-0.011}^{+0.010}}$ & $-1.026_{-0.056}^{+0.062}$ \\
$a_{1}$ & $0.924_{-0.015}^{+0.016}$ & $1.290_{-0.298}^{+0.258}$ \\
$a_{2}$ & $-1.541_{+0.015}^{-0.016}$ & $-1.216_{+2.208}^{+0.216}$ \\
$a_{3}$ & $0.897_{-0.015}^{+0.016}$ & $0.210_{+0.191}^{-0.223}$ \\
\bottomrule\bottomrule
\end{tabularx}
\end{center}\caption{Best-fit parameter values for the pion $\pi^+$ FFs of the {\tt JAM20} at the initial scale $\mu_F=1.3\,{\rm GeV}$, where the parametrization functional form is shown in \eqref{eq:FFpara}, and the errors are the uncertainties calculated by the replica method. The upper (lower) part of the numbers is the difference from the central values corresponding to the upper (lower) bound of the FFs. }\label{tab:FFs}
\end{table}

\bibliographystyle{JHEP}
\bibliography{jet.bib}

\end{document}